\documentclass[aip,preprint]{revtex4-1}

\draft

\usepackage{graphicx}
\usepackage{bm}
\usepackage{color}

\usepackage{float}

\begin{document}

\title{High-mobility p-channel wide bandgap transistors based on $h$-BN/diamond heterostructures}

\author{Yosuke Sasama}
\affiliation{International Center for Materials Nanoarchitectonics (WPI-MANA), \mbox{National Institute for Materials Science, Tsukuba 305-0044, Japan}}
\affiliation{University of Tsukuba, Tsukuba, 305-8571, Japan}

\author{Taisuke Kageura}
\affiliation{International Center for Materials Nanoarchitectonics (WPI-MANA), \mbox{National Institute for Materials Science, Tsukuba 305-0044, Japan}}

\author{Masataka Imura}
\affiliation{Research Center for Functional Materials, National Institute for Materials Science, Tsukuba 305-0044, Japan}

\author{Kenji Watanabe}
\affiliation{Research Center for Functional Materials, National Institute for Materials Science, Tsukuba 305-0044, Japan}

\author{\hspace{12mm}Takashi Taniguchi}
\affiliation{International Center for Materials Nanoarchitectonics (WPI-MANA), \mbox{National Institute for Materials Science, Tsukuba 305-0044, Japan}}

\author{Takashi Uchihashi}
\affiliation{International Center for Materials Nanoarchitectonics (WPI-MANA), \mbox{National Institute for Materials Science, Tsukuba 305-0044, Japan}}

\author{Yamaguchi Takahide}
\affiliation{International Center for Materials Nanoarchitectonics (WPI-MANA), \mbox{National Institute for Materials Science, Tsukuba 305-0044, Japan}}
\affiliation{University of Tsukuba, Tsukuba, 305-8571, Japan}

\begin{abstract}
Field-effect transistors made of wide-bandgap semiconductors can operate at high voltages, temperatures and frequencies with low energy losses, and have been of increasing importance in power and high-frequency electronics. However, the poor performance of p-channel transistors compared with that of n-channel transistors has constrained the production of energy-efficient complimentary circuits with integrated n- and p-channel transistors. The p-type surface conductivity of hydrogen-terminated diamond offers great potential for solving this problem, but surface transfer doping, which is commonly believed to be essential for generating the conductivity, limits the performance of transistors made of hydrogen-terminated diamond because it requires the presence of ionized surface acceptors, which cause hole scattering. Here, we report on fabrication of a p-channel wide-bandgap heterojunction field-effect transistor consisting of a hydrogen-terminated diamond channel and hexagonal boron nitride ($h$-BN) gate insulator, without relying on surface transfer doping. Despite its reduced density of surface acceptors, the transistor has the lowest sheet resistance ($1.4$ k$\Omega$) and largest on-current ($1600$ $\mu$m mA mm$^{-1}$) among p-channel wide-bandgap transistors, owing to the highest hole mobility (room-temperature Hall mobility: $680$ cm$^2$V$^{-1}$s$^{-1}$). Importantly, the transistor also shows normally-off behavior, with a high on/off ratio exceeding $10^8$. These characteristics are suited for low-loss switching and can be explained on the basis of standard transport and transistor models. This new approach to making diamond transistors paves the way to future wide-bandgap semiconductor electronics.
\end{abstract}

\maketitle

Wide bandgap semiconductors, such as silicon carbide (SiC) and gallium nitride (GaN), have significant advantages over silicon for power and high-frequency applications. The high breakdown electric field of wide-bandgap semiconductors can reduce the length of the drift region and increase its doping concentration in field-effect transistors (FETs), leading to reduced on-resistance and conduction loss. The wide bandgap also allows high-temperature operation, thereby reducing the size of cooling systems. SiC-based FETs are used in compact and low-loss power conversion systems because of these benefits. Meanwhile, the high carrier density and high carrier mobility in GaN-based n-channel FETs (high electron mobility transistors; HEMTs) enable these devices to handle high currents and operate at high frequency, making them suitable for RF power applications.

However, the poor performance of p-channel wide-bandgap FETs compared with that of their n-channel counterparts remains an issue. This problem has constrained the production of complementary circuits with integrated n- and p-channel transistors that gain advantages such as low power consumption and simple driving circuitry. \cite{Bad20} The poor performance of p-channel wide-bandgap FETs stems from low hole mobility. This low mobility causes high on-resistance and high conduction loss. It also limits the current handling capability and degrades the high-frequency operation of FETs. Although electron mobility in GaN HEMTs is $1500 - 2000$ cm$^2$V$^{-1}$s$^{-1}$, hole mobility in GaN-based p-channel FETs is below $30$ cm$^2$V$^{-1}$s$^{-1}$. \cite{Bad20,Reu14,Nak10,Cha19,Kri19,Zhe20,Raj20,Kri20} For SiC-based FETs, electron mobility is $30-200$ cm$^2$V$^{-1}$s$^{-1}$, \cite{Nan13,Hat17} while hole mobility is lower than $17$ cm$^2$V$^{-1}$s$^{-1}$. \cite{Bad20,Oka06,Nob09} Even producing any amount of p-type conductivity is a challenge for gallium oxide (Ga$_2$O$_3$). \cite{Pea18}

Diamond is a unique wide-bandgap material that has the potential to solve the above problem. In particular, bulk diamond has the phonon-limited intrinsic hole mobility higher than $2000$ cm$^2$V$^{-1}$s$^{-1}$, \cite{Isb02,Per10,Aki14,Koi18} which is promising for production of high-mobility p-channel transistors. The high intrinsic mobility of diamond relates to its hardness; owing to the high optical phonon frequency and high acoustic phonon velocity of diamond, the number of phonons that participate in carrier scattering is small. \cite{Per10,Aki14} Other outstanding characteristics, such as a wide bandgap ($5.47$ eV), high breakdown field (${\textgreater} 10$ MV cm$^{-1}$), and high thermal conductivity (${\textgreater} 2200$ W m$^{-1}$K$^{-1}$), which are superior to those of SiC, GaN, and Ga$_2$O$_3$, also make diamond attractive for power applications. \cite{Gei18,Don19} Many of the diamond FETs reported to date have used hydrogen-terminated diamond surfaces as a channel. This is because p-type conductivity can be induced by surface transfer doping \cite{Mai00,Str04} without relying on bulk dopants; electron-acceptor materials covering a hydrogen-terminated diamond surface, such as atmospheric adsorbates, acid gases, or high work-function oxides, capture electrons from the diamond valence bands and thereby create holes at the diamond surface. Because of the surface-transfer-doping concept, a common strategy for making hydrogen-terminated diamond FETs has been to coat the diamond surface with such an acceptor material \cite{Gei18,Ver162,Kaw17,Ren17,Kas17,Yu18,Yin18,Gei20} (see Supplementary Section A for details). However, this method is accompanied by negatively charged acceptors near the diamond surface, which act as sources of carrier scattering and reduce carrier mobility. \cite{Li18,Sas20} The negative charges also cause a positive shift in the threshold voltage, leading to normally-on operation, \cite{Ver162,Ren17,Kas17,Yin18,Gei20} \textit{i.e.} a finite channel conductance at zero gate bias. Normally-on operation is undesirable from the viewpoint of fail-safe operation and driving circuit simplicity, particularly in power electronics applications.

\begin{figure}
\includegraphics[width=16truecm]{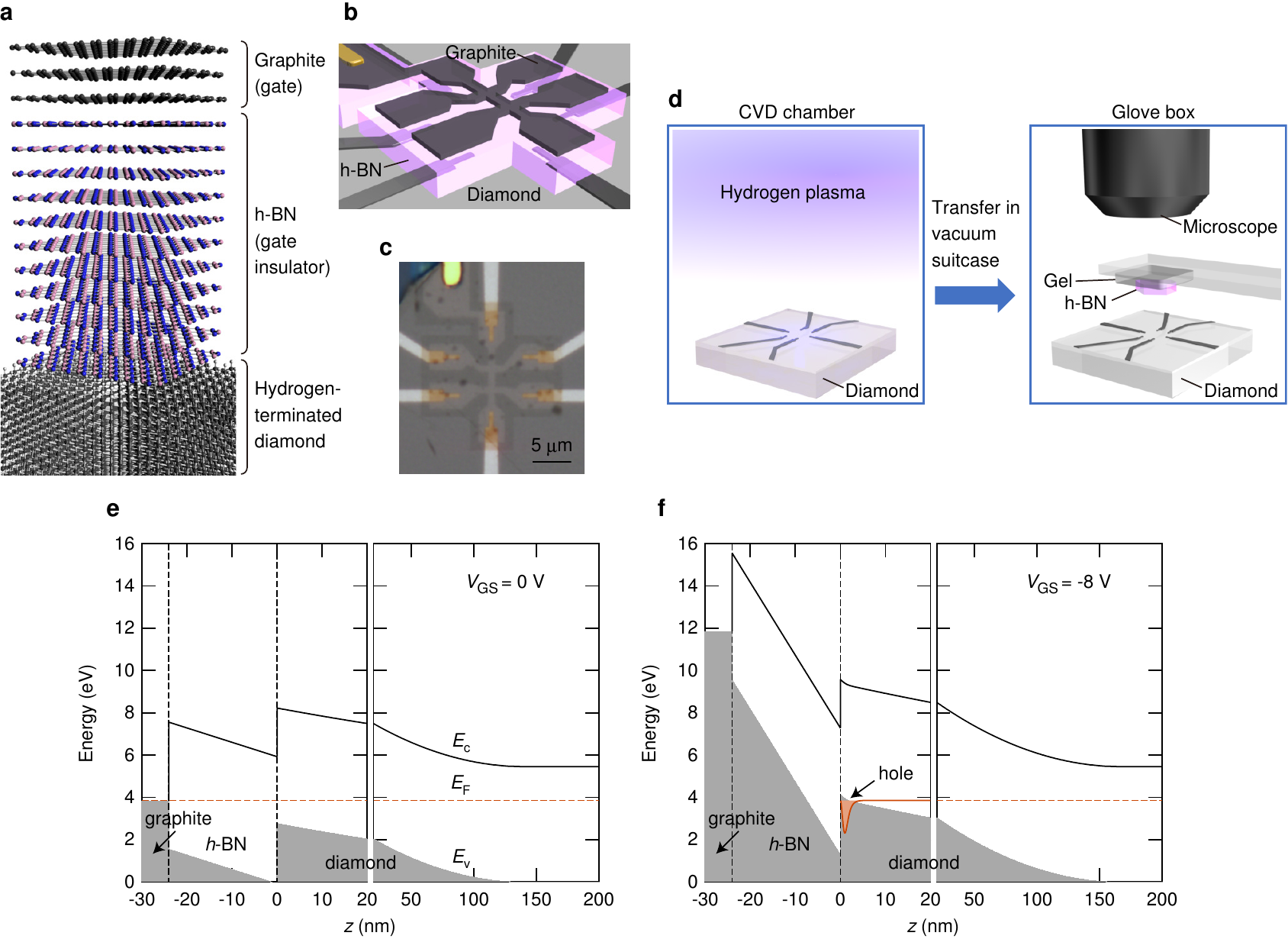}
\caption{\textbf{Diamond field-effect transistor (FET) with a hexagonal boron nitride ($h$-BN) gate insulator and graphite gate.} \textbf{a}, Schematic diagram of heterostructure consisting of graphite/$h$-BN/hydrogen-terminated diamond. $h$-BN is thicker in actual devices (24 nm, corresponding to 80 layers). \textbf{b}, Schematic diagram of diamond field-effect transistor with $h$-BN gate insulator and graphite gate. The diamond surface in the region covered with $h$-BN is hydrogen-terminated. The diamond surface in the other region is oxygen-terminated, which makes the surface highly insulating and provides electrical isolation. \textbf{c}, Optical microscope image of fabricated device. \textbf{d}, Schematic diagram of transfer of diamond from the CVD chamber, where the diamond surface is hydrogenated by hydrogen plasma, to a glove box, where the diamond is laminated with an $h$-BN thin crystal. The transfer is made in a vacuum suitcase without any exposure of the diamond surface to air. \textbf{e,f}, Energy-band diagrams of the graphite (gate; $z{\textless}-24$ nm)/$h$-BN ($-24$ nm ${\textless} z {\textless} 0$)/diamond ($z{\textgreater}0$) heterostructure for applied gate voltages of 0 (\textbf{e}) and -8 V (\textbf{f}). The energy is relative to the valence band maximum deep inside the diamond. An inversion layer of holes is formed at the diamond surface for $V_\mathrm{GS} = -8$ V.}
\end{figure}

In this study, we fabricated high-mobility p-channel wide-bandgap transistors by reducing the density of surface acceptors on hydrogen-terminated diamond. Notably, the transistors show normally-off behavior with high on/off ratios. The presence of surface acceptor states had been believed to be essential for the surface conductivity of hydrogen-terminated diamond, but this study demonstrates that they are unnecessary and even should be removed to improve device performance. The new strategy for fabricating diamond transistors demonstrated in this study opens a pathway to the development of high-performance wide-bandgap p-channel transistors and complementary circuits.

\subsection*{Field-effect transistors based on $h$-BN/diamond heterostructures\\}

Schematic drawings of the diamond FETs fabricated in this study are shown in Figs. 1a and 1b. A gated Hall-bar structure was used to evaluate the carrier density and mobility through Hall measurements (Figs. 1b and 1c). The gate insulator was produced by laminating a hydrogen-terminated diamond surface with a cleaved $h$-BN single crystal, as in our previous studies. \cite{Sas18,Sas19} This method is useful for avoiding the formation of defects in the gate insulator, which occurs in conventional deposition techniques and can cause acceptor states. As a further step in this study, the $h$-BN lamination was conducted without exposing the hydrogen-terminated diamond surface to air (Fig. 1d), which reduced the density of atmospheric acceptors substantially. Furthermore, a thin graphite crystal was used as a gate electrode to reduce disorder at the interface between the gate and the gate insulator. \cite{Zib17} Energy-band diagrams (Figs. 1e and 1f; see Supplementary Section D for details) obtained from a self-consistent solution of Schr\"{o}dinger and Poisson equations illustrate that an inversion layer of holes forms at the diamond surface under an applied gate voltage, without any surface transfer doping. A brief outline of the device fabrication is as follows (the details are given in Methods and Supplementary Fig. S3). After deposition of Ti/Pt and annealing for forming ohmic contacts, the diamond surface was hydrogenated in a chemical vapor deposition (CVD) chamber. The diamond was directly transferred in a vacuum suitcase from the CVD chamber to an Ar-filled glove box, where the diamond was laminated with $h$-BN and graphite by using the Scotch-tape-exfoliation and dry-transfer techniques. \cite{Cas14} The subsequent processes were carried out outside the glove box, including etching of the graphite and $h$-BN into a Hall-bar shape and deposition of Ti/Au for leads and bonding pads.

\begin{figure}
\includegraphics[width=7.5truecm]{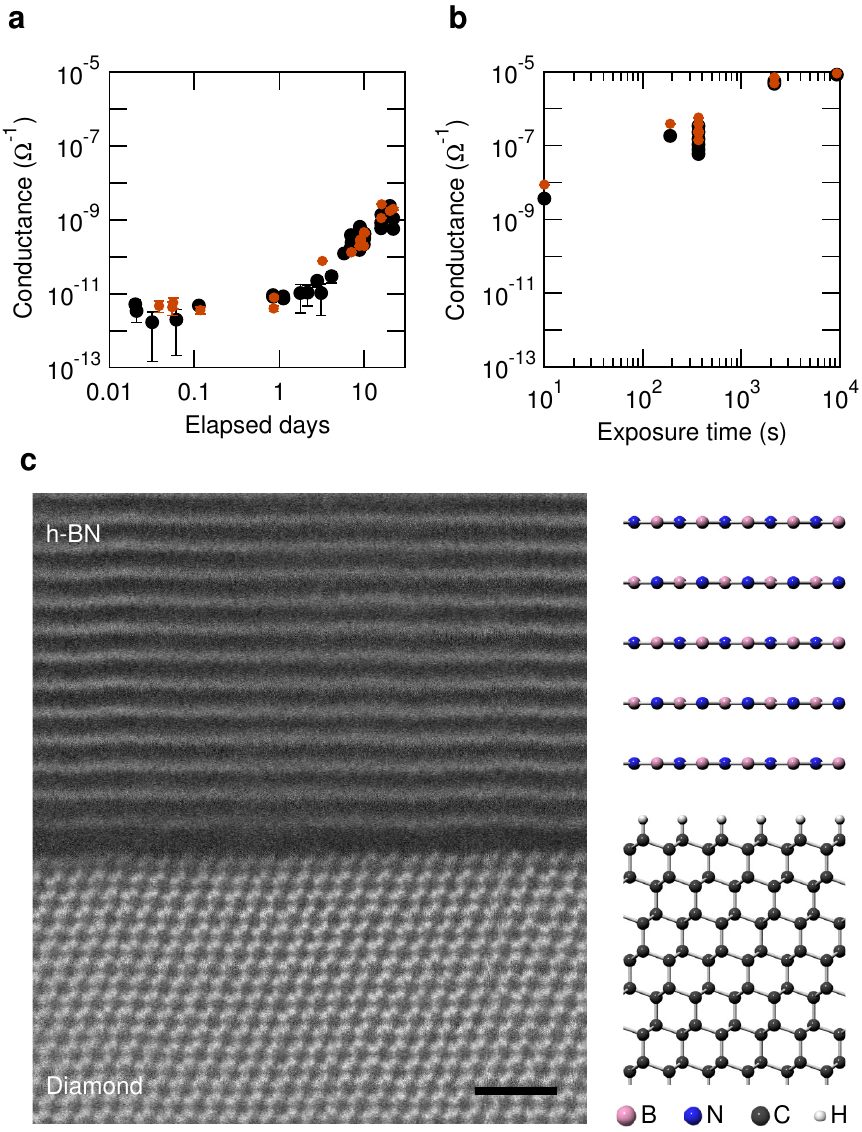}
\caption{\textbf{Reduction of acceptor density on hydrogen-terminated diamond with no exposure to air.} \textbf{a}, Surface conductivity of hydrogen-terminated diamond kept in the glove box after vacuum transfer from CVD chamber. Black and brown dots represent two different diagonal positions on the diamond substrate at which prober needles are contacted. The error bars represent the standard error from linear fits of the current-voltage characteristics. \textbf{b}, Surface conductivity vs. total air-exposure time. After measurements in which the diamond was kept in the glove box for about twenty days, as shown in (\textbf{a}), a test was conducted to see the effect of exposure to air. The diamond was exposed to air for a certain time and brought back to the glove box, where the surface conductivity was measured. This procedure was repeated while increasing the exposure time. The surface conductivity is plotted against cumulative total time. \textbf{c}, Scanning transmission electron microscope (STEM) image of $h$-BN/hydrogen-terminated (111) diamond heterostructure prepared with the air-free process. The image was taken from the $[\bar{1} 1 0]$ crystallographic direction of diamond. Hydrogen atoms are invisible in the STEM image. The scale bar is 1 nm.}
\end{figure}

The reduction of the density of atmospheric acceptors in the above fabrication process was confirmed by measuring the surface conductivity of hydrogen-terminated diamond substrates (not laminated with $h$-BN) by using an electrical prober system placed in the glove box. The surface conductivity was kept low when the diamond had been transferred from the CVD chamber and stored in the glove box. The resistance was kept higher than ${\approx}10^{11}$ $\Omega$ for at least three days (Fig. 2a). Note that the $h$-BN lamination for fabricating FETs was completed within 3 hours of the diamond transfer from the CVD chamber to the glove box. The gradual increase in the conductivity may be due to adsorption of remaining atmospheric gases in the glove box. When the diamond surface was exposed to air, it turned into a conductive state with a resistance of $\approx 100$ k$\Omega$, which is as expected because hydrated ions in atmospheric adsorbates act as surface acceptors \cite{Mai00,Str04} (Fig. 2b). The increased conductivity rules out the possibility that the low conductivity shown in Fig. 2a was caused by failure of the hydrogenation of the diamond surface. These observations indicate that the atmospheric acceptors that cause the surface conductivity can be substantially reduced by using the above fabrication process. This was further evidenced by the normally-off behavior of the FETs described below. A scanning transmission electron microscope (STEM) image of an $h$-BN/diamond heterostructure prepared with the air-free process (Fig. 2c) indicates that the distance between the first $h$-BN layer and the hydrogen-terminated diamond surface is nearly the same as the interlayer spacing (0.333 nm) of $h$-BN. (Hydrogen atoms are invisible in the STEM image. The C-H bond length is estimated to be 0.112 nm for hydrogen-terminated (111) diamond surfaces. \cite{Squ06}) The STEM image of a larger-area $h$-BN/diamond heterostructure (Supplementary Fig. S5a) shows that the spacing between the first $h$-BN layer and diamond is uniform over a large area. These images also suggest a low density of contaminants at the interface.

\begin{figure}[H]
\includegraphics[width=16truecm]{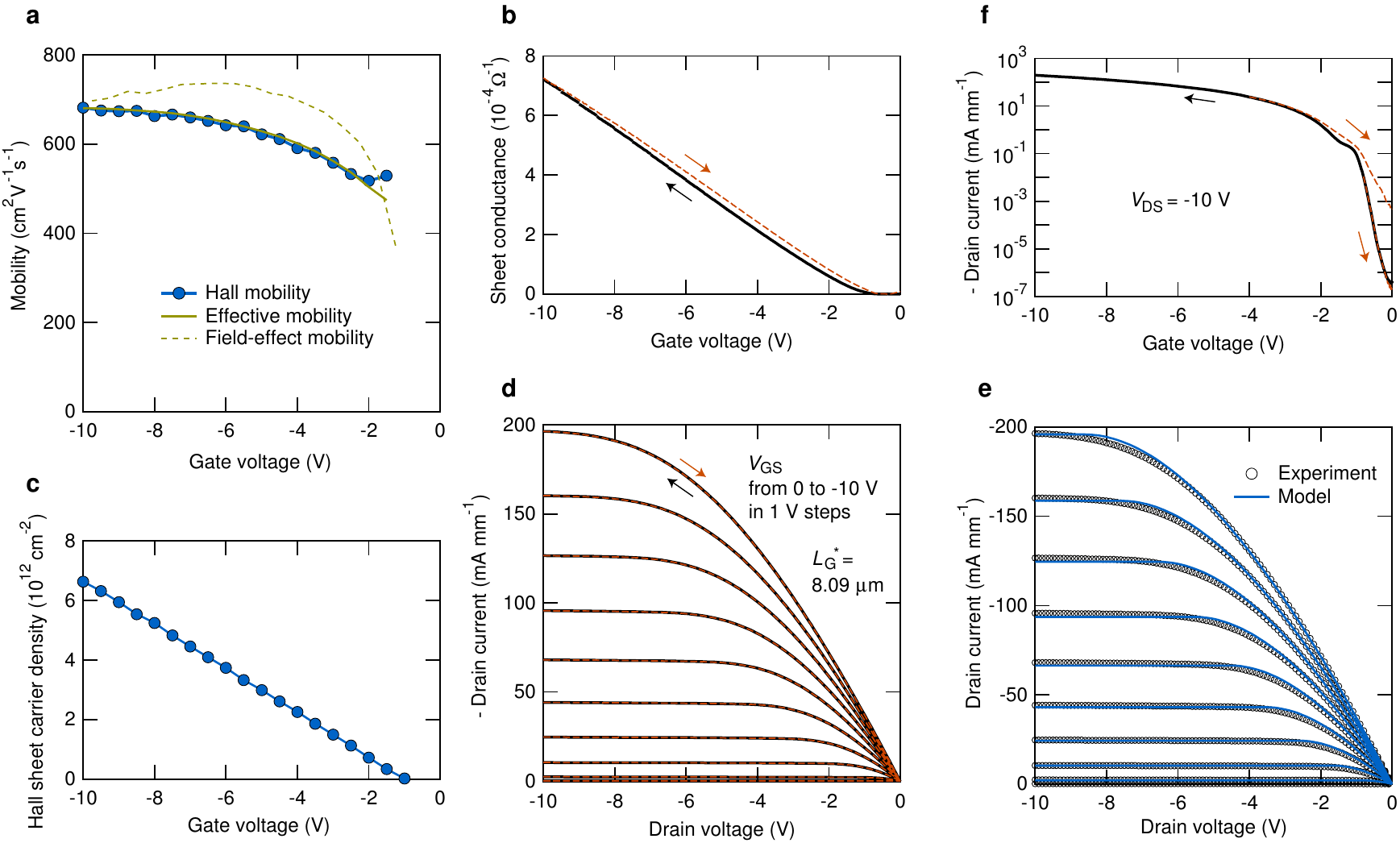}
\caption{\textbf{Electrical characteristics of diamond FET at room temperature.} \textbf{a}, Hall, effective, and field-effect mobilities as a function of gate voltage. A dielectric constant value of $3.2$ (Refs. \cite{Ste20,You12}) for h-BN was used for calculating the effective and field-effect mobilities. \textbf{b}, Transfer characteristics in the linear region. The channel sheet conductance was measured with the four-probe configuration. The solid and dashed lines were obtained while the gate voltage was swept from 0 to -10 V and from -10 to 0 V, respectively. The hysteresis, in which the magnitude increased with maximum applied gate voltage, may be due to relocation of residual surface acceptors under electric fields. \textbf{c}, Hall carrier density as a function of gate voltage. \textbf{d}, Output characteristics. The drain current measured with the two-probe configuration is plotted as a function of the drain voltage for different gate voltages. \textbf{e}, Measured output characteristics (circles, same as black lines in \textbf{d}) and calculated ones (lines) using a charge-sheet model (see Supplementary Section E for the detailed modeling). \textbf{f}, Drain current as a function of gate voltage at a drain voltage of -10 V. The solid line corresponds to a gate sweep from 0 to -10 V. The two dashed lines correspond to gate sweeps from -0.8 to 0 V and -4 to 0 V.}
\end{figure}

\newpage

\subsection*{Room-temperature electrical characteristics of $h$-BN/diamond FETs\\}

The electrical properties of our FETs were evaluated by making conductivity and Hall-effect measurements. Figure 3a shows the gate voltage dependence of Hall mobility at room temperature. The sign of the Hall voltage is positive, indicating that the charge carriers are holes. The Hall mobility reaches $680$ cm$^2$V$^{-1}$s$^{-1}$ at a gate voltage $V_\mathrm{GS}$ of $-10$ V. This is higher than those of FETs fabricated with air-exposed hydrogen-terminated diamond and $h$-BN\cite{Sas18}, indicating that the reduction in surface-acceptor density is effective in improving mobility. Moreover, this room-temperature mobility is the highest ever reported for p-channel FETs based on a diamond surface channel or any other wide-bandgap semiconductor. The effective and field-effect mobilities, $\mu_{\mathrm{eff}}= \frac{t_{\mathrm{hBN}}}{\epsilon _{\mathrm{hBN}}}\frac{\sigma}{\left|V_{\mathrm{GS}} -V_{\mathrm{th}}\right|}$ and $\mu_{\mathrm{FE}}= \frac{t_{\mathrm{hBN}}}{\epsilon _{\mathrm{hBN}}}\left|\frac{\partial \sigma}{\partial V_{\mathrm{GS}}}\right|$ ($t_{\mathrm{hBN}}$ and $\epsilon _{\mathrm{hBN}}$, thickness and dielectric constant of $h$-BN; $\sigma$, sheet conductance in the linear regime; $V_{\mathrm{th}}$, threshold voltage), were also evaluated from the transfer curve (Fig. 3b). If we use $\epsilon_\mathrm{hBN}/\epsilon_0 = 3.2$, \cite{Ste20,You12} the gate voltage dependence of the effective mobility is in good agreement with that of the Hall mobility, as shown in Fig. 3a. The field-effect mobility is larger than the effective mobility especially at small gate voltages. This is because $\mu_{\mathrm{FE}}= {\mu}_{\mathrm{eff}} + (V_{\mathrm{GS}}-V_{\mathrm{th}})\frac{\partial {\mu}_{\mathrm{eff}}}{\partial V_{\mathrm{GS}}}$ and the sign of $(V_{\mathrm{GS}}-V_{\mathrm{th}})\frac{\partial {\mu}_{\mathrm{eff}}}{\partial V_{\mathrm{GS}}}$ is positive (Supplementary Section F).

Normally-off operation of our FET was clearly indicated by the transfer curve and the gate voltage ($V_\mathrm{GS}$) dependence of the Hall carrier density ($p_\mathrm{Hall}$) at room temperature (Fig. 3c). A linear fit to the $p_\mathrm{Hall}-V_\mathrm{GS}$ curve gives a threshold voltage $V_\mathrm{th}$ of $-0.99$ V. This threshold voltage is close to the value predicted by the standard formula using device parameters such as the work function difference between hydrogen-terminated diamond and graphite, as described below. The carrier density reaches $6.6\times10^{12}$ cm$^{-2}$ at $V_\mathrm{GS} = -10$ V. At this gate voltage, the sheet conductance is $7.2\times10^{-4}$ $\Omega^{-1}$, which is equivalent to a sheet resistance of 1.4 k$\Omega$. This on-sheet-resistance is the lowest ever reported for diamond field-effect transistors. Since on-resistance is usually sacrificed for normally-off operation, \cite{Kit17,JFZha20,Liu17,Oi19} the low on-resistance compatible with normally-off operation in our FET is remarkable. This resistance is even comparable to the lowest sheet resistances reported for hydrogen-terminated diamond exposed to NO$_2$ or coated with WO$_3$ or V$_2$O$_5$ with no FET structure. \cite{Kas17,Tor17,Ver18} The low sheet resistance of NO$_2$-exposed or WO$_3$- or V$_2$O$_5$-coated hydrogen-terminated diamond is due to a high carrier density on the order of $10^{14}$ cm$^{-2}$, whereas the low sheet resistance of our FET is mainly due to the high carrier mobility.

The output characteristics (Fig. 3d) also indicate the excellent performance of our FET. The output curves display nearly perfect saturation in drain current. The output currents, as shown in Fig. 3e, can be quantitatively explained on the basis of a long-channel transistor model (see Supplementary Section E for the detailed modelling). The maximum current normalized by the channel width ($I_\mathrm{D}^\mathrm{max}/W_\mathrm{G}$) is $200$ mA mm$^{-1}$. This value is again exceptionally high among those of normally-off diamond FETs, \cite{JFZha20,Wan18,Ren18,Fei20} despite the long gate length $L_\mathrm{G}^{*}$ of 8.09 $\mu$m. (See Supplementary Table S1 for the device dimensions.) Note that the maximum current is generally inversely proportional to the gate length. The gate-length-normalized maximum current ($L_\mathrm{G}^{*}I_\mathrm{D}^\mathrm{max}/W_\mathrm{G}$) is $1600$ $\mu$m mA mm$^{-1}$, which is the highest yet reported for p-channel FETs based on diamond or any other wide-bandgap semiconductor. The transconductance is also high, reaching 37 mS mm$^{-1}$ (Supplementary Fig. S6a). These high on-state characteristics are the result of the high carrier mobility achieved by the reduced surface-acceptor density.

Despite the high on-state characteristics, the FETs can be switched off with a low source-drain leakage current. The gate voltage dependence of the drain current in the saturation region (drain voltage $V_\mathrm{DS} = -10$ V) gives an on-off ratio larger than $10^8$ (Fig. 3f). The leakage current in the off state ($V_\mathrm{GS} = 0$ V) is almost at the noise level of our measurement setup. The drain current shows hysteresis and its magnitude increases with maximum applied gate voltage. The hysteresis may be caused by relocation of the residual surface acceptors under electric fields. The minimum subthreshold swing $SS$ is $130$ mV dec$^{-1}$ at $V_\mathrm{GS} = -0.55$ V. The interface trap density $D_\mathrm{it}$ can be evaluated using the equation, \cite{Sze07}
\begin{eqnarray}
 SS = \ln{10}\frac{dV_{\mathrm{GS}}}{d(\ln{I_{\mathrm{D}}})} = \ln{10}\frac{k_{B}T}{e}\frac{C_{\mathrm{hBN}} + C_{\mathrm{depl}} +{e}^{2}D_{\mathrm{it}}}{C_{\mathrm{hBN}}},
 \label{eq:SSDit}
\end{eqnarray}where $k_B$ is the Boltzmann constant, $T$ the absolute temperature, ${e}$ the elementary charge, $C_\mathrm{hBN}$ ($=\epsilon_\mathrm{hBN}/t_\mathrm{hBN}$) the capacitance of the $h$-BN gate insulator, and $C_\mathrm{depl}$ the capacitance of the diamond depletion layer. If we use $t_\mathrm{hBN}=24$ nm and $\epsilon_\mathrm{hBN}/\epsilon_0= 3.2$, $C_\mathrm{hBN}=0.12$ $\mu$Fcm$^{-2}$. $C_\mathrm{depl}$ is estimated to be $3.0\times10^{-2}$ $\mu$Fcm$^{-2}$ from the calculated depletion layer thickness for $N_D = 500$ ppb and $N_A = 7$ ppb, where $N_D$ and $N_A$ are donor (nitrogen) and acceptor (boron) concentrations obtained from a secondary ion mass spectrometry measurement. Using these values, $D_\mathrm{it}$ is evaluated to be $6.8\times10^{11}$ cm$^{-2}$eV$^{-1}$. This interface state density is lower than most of values reported for diamond/insulator interfaces in recent literature. \cite{Sah19,Mat19, Zha202} This is presumably due to the dangling-bond-free surface of $h$-BN and reduced adsorbate density. The gate leak current is almost at the noise level ($5\times10^{-13}$ A) for gate voltages between 0 and -10 V. This current corresponds to $3\times10^{-7}$ A cm$^{-2}$ if it is normalized by the total area of the graphite gate electrode. If it is normalized by the channel width, it corresponds to $5\times10^{-7}$ mA mm$^{-1}$ (Supplementary Fig. S6b). The gate leak current is sufficiently small in both representations. The gate-source breakdown field is at least higher than 4.2 MV/cm (=10 V/ 24 nm).

\begin{figure}
\includegraphics[width=8truecm]{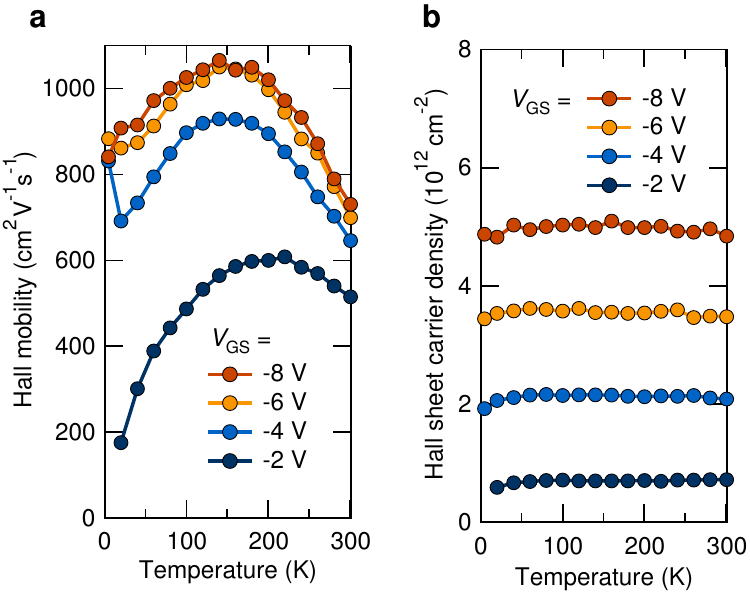}
\caption{\textbf{Temperature-dependent electrical characteristics of diamond FET.} Temperature dependence of Hall mobility (\textbf{a}) and Hall carrier density (\textbf{b}) at different gate voltages.}
\end{figure}

\subsection*{Temperature dependence of electrical properties of $h$-BN/diamond FETs\\}

The temperature dependence of the electrical properties gives another insight into the quality of our FET channel (Fig. 4). Generally, the mobility of the carriers induced by surface acceptors on a hydrogen-terminated surface decreases as the temperature falls below room temperature. \cite{Kas17,Pet20} This is because negative charges near the surface are the dominant source of carrier scattering that determines the mobility. Lowering the temperature increases the population of lower energy holes, which are influenced more significantly by the potential fluctuation. The mobility of our FETs clearly increases with decreasing temperature from $300$ to $150$ K. The mobility exceeds $1000$ cm$^2$V$^{-1}$s$^{-1}$ at $150$ K for $V_\mathrm{GS} = -8$ V (Fig. 4a). The negative sign of the derivative of mobility with respect to temperature indicates the contribution of phonon scattering, which is now clearly observable in the temperature range of $150-300$ K due to the suppression of ionized impurity scattering (see Supplementary Section C and Fig. S12a for the detailed modelling of mobility). In addition, the high mobility persists at low temperature, even for a low carrier density of $2\times10^{12}$ cm$^{-2}$ (Figs. 4a and 4b). This carrier density is lower than those required to maintain high mobility in previous studies. \cite{Sas18,Sas19,Yam14} Furthermore, the carrier density in Fig. 4b is nearly independent of temperature down to the lowest temperature ($4.5$ K), indicating that freezing-out of carriers does not occur even at such a low temperature. These observations all indicate an improvement in channel quality and demonstrate that surface acceptors are unnecessary for obtaining highly conductive states in hydrogen-terminated diamond. Gate controllability is retained even at cryogenic temperatures (Supplementary Fig. S7), which may be used to fabricate a device that works in an environment with large temperature variations, e.g. in outer space.

\subsection*{Mobility and normally-off operation of $h$-BN/diamond FETs \\}

\begin{figure}
\includegraphics[width=7truecm]{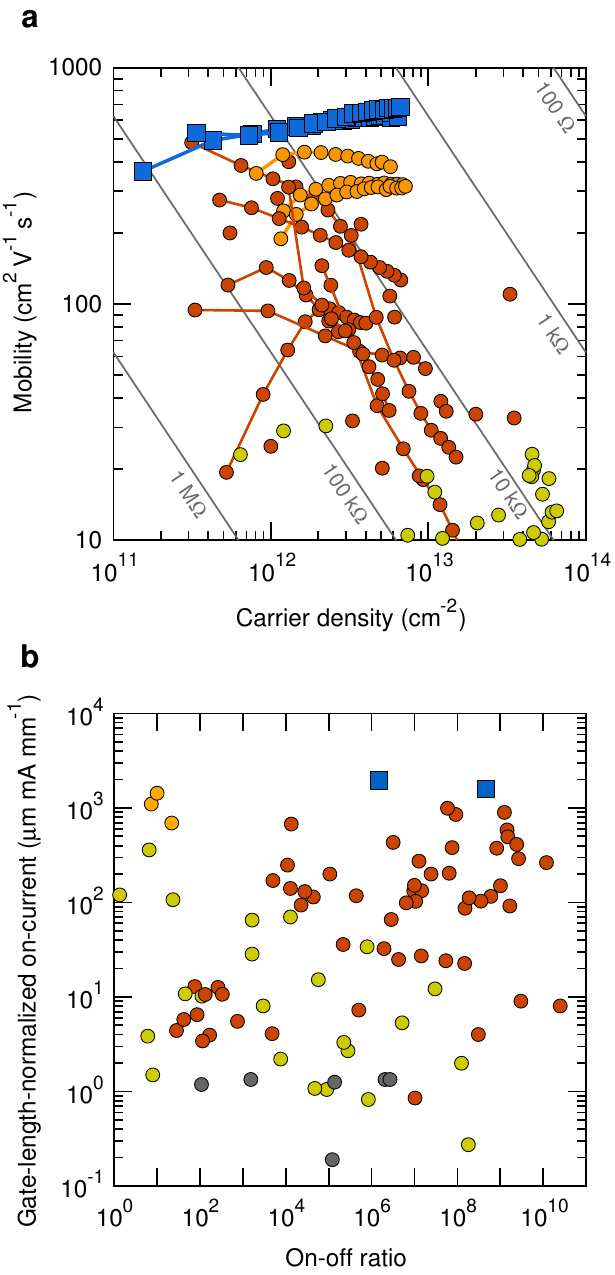}
\caption{\textbf{Comparison of room-temperature performances of diamond FET and prior work.} \textbf{a}, Mobility and carrier density of our present FETs (blue squares; two devices), other diamond FETs (brown and orange dots; orange dots, our previous FETs fabricated with air-exposed hydrogen-terminated diamond and $h$-BN), and p-channel GaN-based heterostructures (green dots). References for the data are given in Supplementary Section B. The mobility-extraction methods for the diamond FETs are tabulated in Supplementary Table S2. Lines indicate the sheet resistances of 100 $\Omega$-1 M$\Omega$. \textbf{b}, Normalized on-current and on-off ratio of our present FETs (blue squares; two devices), other diamond FETs (brown and orange dots; orange dots, our previous FETs fabricated with air-exposed hydrogen-terminated diamond and $h$-BN), p-channel GaN FETs (green dots), and p-channel SiC FETs (grey dots). References for the data are given in Supplementary Section B.}
\end{figure}

Figure 5 benchmarks the performance of our FETs at room temperature against the literature. The mobility is the highest among FETs based on a diamond surface channel reported to date (Fig. 5a). It is also higher than the ${\le} 30$ cm$^2$V$^{-1}$s$^{-1}$ of p-channel GaN-based heterostructures \cite{Reu14,Nak10,Cha19,Kri19,Zhe20,Raj20,Kri20} and $7 - 17$ cm$^2$V$^{-1}$s$^{-1}$ of p-channel SiC-based FETs. \cite{Oka06,Nob09} (The mobilities for SiC are not shown in Fig. 5a because the carrier densities are not reported in \cite{Oka06,Nob09}.) The high mobility lowers the channel sheet resistance, approaching $1$ k$\Omega$, and increases the on-current $I_\mathrm{ON}$ ($=I_\mathrm{D}^\mathrm{max}/W_\mathrm{G}$). The gate-length-normalized on-current $I_\mathrm{ON}L_\mathrm{G}^{*}$ is the highest among p-channel FETs made of diamond and other wide-bandgap semiconductors (Fig. 5b). Importantly, normally-off behavior with a large on-off ratio is obtained simultaneously.

The normally-off behavior of our FETs can be understood by considering that the densities of atmospheric adsorbates and the resulting negative charges are reduced in the FETs. The threshold voltage is given by \cite{Sze07}
\begin{eqnarray}
V_\mathrm{th} = -\frac{{e}(n_\mathrm{depl} - n_\mathrm{ic}) t_\mathrm{hBN}}{\epsilon_\mathrm{hBN}} + \psi_\mathrm{s}(p_\mathrm{2D}\to0) + \phi_\mathrm{ms}.
 \label{eq:Vgs}
\end{eqnarray}
Here, $n_\mathrm{depl}$ is the sheet density of the fixed charge in the depletion layer and $n_\mathrm{ic}$ is the net sheet density of the interface negative charge. $\psi_{\mathrm{s}}(p_\mathrm{2D}\to0)$ (${\textless}0$) is the surface potential in the limit of low hole sheet density $p_\mathrm{2D}$. $e\phi_\mathrm{ms}=e\phi_\mathrm{m}-e\phi_\mathrm{s}$ is the difference between the work function (e$\phi_\mathrm{m}$) of the graphite gate and that (e$\phi_{\mathrm{s}}$) of hydrogen-terminated diamond. This equation indicates that a decrease in the interface negative charge causes a negative shift in the threshold voltage $V_\mathrm{th}$. The threshold voltage calculated using the above equation is in good agreement with the measured threshold voltage (Supplementary Section D and Fig. S12c). The increase in donor (e.g. nitrogen) density in diamond increases $n_\mathrm{depl}$ and causes a negative shift in $V_\mathrm{th}$, but this enhances undesirable ionized impurity scattering. A partial UV ozone treatment \cite{Kit17,JFZha20} achieves normally-off operation presumably by increasing the spatially averaged value of $e\phi_\mathrm{s}$, but non-uniform hydrogen termination should cause local potential fluctuations that result in carrier scattering. Introducing positive charges into the gate insulator to compensate the interface negative charge \cite{Liu17} should also cause local potential fluctuations. By contrast, the normally-off operation in our FETs is based on a reduction in interface negative charge, which decreases carrier scattering and thus enhances mobility. Note that the choice of a gate material with an appropriate work function $e\phi_\mathrm{m}$ would enable the threshold voltage to be tuned without lowering the mobility.

The holes in our FETs are generated in the absence of acceptors by an upward band bending due to gate bias (Fig. 1f), a mechanism equivalent to the formation of the inversion channel in Si metal-oxide-semiconductor (MOS) FETs, where electrons are generated in the absence of donors. In fact, the output characteristics are in excellent agreement with those predicted by the standard model for an inversion channel (Fig. 3e). The positive charge of the holes is mostly balanced by the negative charge of electrons in the gate electrode. This remote and uniform negative charge does not scatter holes, and leads to an increase in mobility. This means of increasing the mobility is analogous to that for GaAs HEMTs, in which a doped layer is spatially separated from the channel of the transistors. Although the doped layer is required to compensate for the high density of surface states in GaAs HEMTs, \cite{Dav98} remote dopants are unnecessary in our FETs, thanks to the low density of surface states in $h$-BN. The reason for using the hydrogen-terminated surface is not the surface transfer doping phenomenon, but its shallow valence bands, which are rooted in the presence of C$^{\delta-}$-H$^{\delta+}$ dipoles, \cite{Squ06} the low density of surface states, and the availability of good ohmic contacts, where this last point should relate to the first two.

Despite the improved transport characteristics, the decrease in mobility at low temperature (Fig. 4a) and the increase in mobility with carrier density (Fig. 5a) suggest that there is still some contribution from charged impurity scattering (see Supplementary Section C and Fig. S12 for the modelling of mobility). The charge that causes the carrier scattering may arise from atmospheric gas that was not sufficiently removed by the purifier of the glove box and adsorbed to the diamond surface. Alternatively, charge may have become trapped in interface states due to defects at the surface of diamond or $h$-BN. The interface state density estimated from the subthreshold swing is $6.8\times10^{11}$ cm$^{-2}$eV$^{-1}$, as described above. A plausible origin of the interface states is defects in diamond or $h$-BN near the interface, such as broken bonds or impurities. The defects may be imperfections of hydrogen termination (e.g., dangling bonds, or a different kind of termination such as oxygen), lattice distortions, or carbon vacancies in diamond. Boron or nitrogen vacancies and carbon impurities may be present in $h$-BN. \cite{JZha20} These defects possibly act as interface states of the donor, acceptor, or amphoteric type depending on their kind. For instance, a C=C bond formed near the surface of diamond has been suggested to act as an acceptor state. \cite{Sta19} A carbon dangling bond (missing H termination) at the diamond surface may act as a donor state near the valence band, like the P$_\mathrm{b}$ center (silicon dangling bond) at the Si/SiO$_2$ interface. \cite{Zha202} Such donor-type interface states should be positively charged and reduce mobility. Acceptor-type interface states should have less of an effect on mobility than donor-type interface states, because they are neutral when the Fermi level is lower than their energy levels, which should be the case particularly when the carrier density is high. Differences in mobility and threshold voltage between the two devices fabricated in this study, which seem to relate to the different environmental conditions for the two devices before the $h$-BN lamination (see Methods), suggest that some adsorbates still exist and are causes of the reductions in both mobility and threshold voltage. Indeed, a detailed analysis indicates that interface negative charges due to remaining adsorbates mainly caused the reduction in mobility, while positive charges of donor-type interface states additionally contributed to it (see Supplementary Section D). Further reduction in atmospheric adsorbates and preparation of an interface with a lower density of structural defects, e.g. using atomically flat diamond, \cite{Tok08,Yam14} will increase mobility.

\subsection*{Conclusions\\}

In summary, high-mobility p-channel wide-bandgap transistors were fabricated by using hydrogen-terminated diamond/$h$-BN heterostructures. It was found that surface acceptors, which had been introduced in previous studies on hydrogen-terminated diamond FETs, are unnecessary for hole transport. In fact, our findings indicate that they should be removed to achieve higher hole mobility, better on-state characteristics and normally-off operation. Reducing defects would also increase device stability and reliability, which are important requirements for practical applications. The FET characteristics were quantitatively explained on the basis of standard FET models, which provides a solid basis for further development of diamond FETs. The room-temperature mobility of $680$ cm$^2$V$^{-1}$s$^{-1}$, sheet resistance of $1.4$ k$\Omega$, and on-current of $1600$ $\mu$m mA mm$^{-1}$ of our FET are the best values reported for p-channel FETs made of wide-bandgap semiconductors. These high values illustrate the great potential of diamond FETs for power and high-frequency electronics applications. Particularly promising would be a complementary circuit in which a p-channel diamond FET and an n-channel GaN-based FET are integrated. \cite{Bad20,Kaw17} As well, high-frequency high-output diamond FETs with an intrinsic diamond heat spreader could be fabricated, e.g., by using the heterostructure shown in this study in the channel region while using surface transfer doping in the access region. The hole mobility in our FET increases with decreasing temperature from $300$ to $150$ K and exceeds $1000$ cm$^2$V$^{-1}$s$^{-1}$ at $150$ K. The contribution of phonon scattering to mobility in this temperature range, which had been masked by ionized impurity scattering in previous studies, is now clear. Since the phonon-limited hole mobility is higher than $2000$ cm$^2$V$^{-1}$s$^{-1}$ at room temperature in bulk diamond, \cite{Isb02,Per10,Aki14,Koi18} there should still be room for improving the mobility of diamond FETs. \cite{Sas20,Dal20} This can be achieved by further reducing extrinsic and intrinsic defects near the diamond/gate insulator interface.

\subsection*{Methods}
\subsection*{Device fabrication.}

Field-effect transistors were fabricated on IIa-type (111) single-crystalline diamonds synthesized in a high-temperature, high-pressure process (purchased from Technological Institute for Superhard and Novel Carbon Materials; TISNCM). The (111) orientation was chosen because of its tendency to have higher surface conductivity than the (001) orientation\cite{Kaw12,Yam13} and the small lattice mismatch between $h$-BN and hydrogen-terminated (111) diamond surface.\cite{Sas18} Experimental results for two devices (C1 and C2) are described in this paper. The size of the diamonds was $2\times2\times0.4$ mm for device C1 and $2.5\times2.5\times0.3$ mm for device C2. A secondary ion mass spectrometry measurement on a purchased diamond substrate with the same quality specifications indicated that the concentration of nitrogen was 0.5 ppm ($8\times10^{16}$ cm$^{-3}$) and that of boron was 7 ppb ($1\times10^{15}$ cm$^{-3}$). The surfaces of the diamonds were polished and cleaned in hydrofluoric acid at room temperature and in a mixture of sulfuric and nitric acid at 200\r{}C before device fabrication. Hall-bar electrodes were produced using electron-beam lithography and deposition of Ti/Pt ($5$ nm / $5$ nm) (Supplementary Fig. S3a). The diamond was annealed in H$_2$ gas at a temperature of $650$\r{}C for $35$ min and exposed to hydrogen plasma at $600$\r{}C for $10-12$ min in a microwave-plasma-assisted CVD chamber (Seki Technotron Corporation, AX5200-S). The diamond was then annealed in H$_2$ gas at $650-710$\r{}C for $35$ min and exposed to hydrogen plasma at $600-670$\r{}C for $10$ min in another CVD chamber (Seki Technotron Corporation, AX5000). The flow rate of the H$_2$ gas was $500$ standard cubic centimeters per minute (sccm). These processes resulted in the formation of TiC ohmic contacts and hydrogenation of the diamond surface (Fig. S3b). The diamond was transferred from the second CVD chamber to an Ar-filled glove box in a custom-made vacuum suitcase without exposing the diamond surface to air. Within $30$ hours before the diamond transfer, single-crystalline $h$-BN was cleaved in the glove box using Scotch tape and transferred onto a PDMS (polydimetylsiloxane) sheet (Gel-Pak, PF-60-X4) on an acrylic plate. The diamond surface was laminated with the cleaved $h$-BN thin crystal under an optical microscope in the glove box by using a dry transfer technique \cite{Cas14} (Fig. S3c). The lamination was completed within 3 hours after the diamond transfer from the CVD chamber to the glove box. The sample was then annealed at 300\r{}C in Ar in the glove box for three hours (Fig. S3d). A graphite crystal (Kish graphite, purchased from CoorsTek) for the gate electrode was transferred onto the $h$-BN in a similar manner, followed by annealing at 300\r{}C for an hour (Figs. S3e and S3f). The sample was then taken out of the glove box. The thickness of the h-BN and graphite was measured using atomic force microscopes (Hitachi High-Tech Corporation, AFM5100N, for device C1, and L-trace II, for device C2) in a tapping mode, by using a silicon tip. The thickness of the h-BN was $24$ nm for devices C1 and C2 (Fig. S2). The thickness of the graphite was $6$ nm for device C1 and $20$ nm for device C2. The graphite and $h$-BN were etched into a Hall-bar shape using plasma generated from a mixture of N$_2$, O$_2$, and CHF$_3$ gases with flow rates of $96$ sccm, $2$ sccm, and $2$ sccm at the total pressure of $10$ Pa (Figs. S3g and S3h). The etch rates of the graphite and h-BN were $\ge$7 and $\ge$16 nm/min for an RF power of $35$ W. This process also converted the diamond surface, except the region under the $h$-BN, into an oxygen-terminated one, which was used for device isolation. Another flake of $h$-BN was placed (Fig. S3i) so that the gate lead deposited on it did not touch the hydrogen-terminated surface at the edge of the etched $h$-BN, which could otherwise have resulted in gate leak. The thickness of the h-BN was ${\approx} 73$ nm for device C1 and ${\approx} 63$ nm for device C2. Finally, Ti/Au ($10$ nm / $100$ nm) was deposited for leads and bonding pads (Fig. S3j).

The vacuum suitcase used for fabrication of device C1 was equipped with a non-evaporated getter pump and the pressure was lower than ${10}\time10^{-7}$ Torr during the transfer. The diamonds for device C2 and the STEM specimen were transferred using the same suitcase, but in this case, it was just vacuum-sealed with no pump used during the transfer. The pressure during the transfer was lower than $3{\times}10^{-5}$ Torr in this case. The Ar gas in the glove box was circulated via a purifier consisting of molecular sieves and heated copper to maintain the purity of the Ar gas with the oxygen content below $0.5$ ppm and water content below $2$ ppm. The fabrication of device C1 used another external gas purifier (NuPure Optics, Model 4000 OA), which could remove acid gases and was inserted between the outlet of the purifier described above and the inlet of the glove box. The supplemental gas for controlling the inner pressure of the glove box was argon with a purity of $99.9999$\% provided from a cylinder.

\subsection*{Measurement setup.}

The devices were set in a cryogen-free cryostat with a superconducting magnet. The transfer and Hall characteristics were measured with a gated Hall-bar configuration using a source-measure unit (Keysight Technologies, B2901A) for biasing the gate, a function generator (Agilent Technologies, 33220A) for biasing the drain, and two voltage amplifiers (DL Instruments, 1201) for measuring the longitudinal and Hall voltages, and a current preamplifier (DL Instruments, 1211) for measuring the drain current. The measurements were performed by feeding a small positive and negative current, $I_{D+}$ and $I_{D-}$, in the linear regime (by applying a positive and negative voltage through a $10$ M$\Omega$ series resistor). The absolute value of the applied voltage was smaller than $300$ mV; therefore, the drain current was less than $30$ nA, corresponding to $\le(3.4-3.5)\times10^{-2}$ mA mm$^{-1}$. The sheet resistance $\rho$ was calculated as $\rho=(V_\mathrm{p+}-V_\mathrm{p-})/(I_\mathrm{D+}-I_\mathrm{D-})(W_\mathrm{G}/L_\mathrm{p})$. Here, $V_\mathrm{p+}$ and $V_\mathrm{p-}$ are the voltage between the longitudinal voltage probes of the Hall bar for the positive and negative currents, $L_\mathrm{p}$ is the distance between the voltage probes, and $W_\mathrm{G}$ is the gate width. The sheet conductance $\sigma$ was obtained as $1/\rho$. A magnetic field sweep between $-1$ and $1$ T was used for the Hall measurements. The Hall coefficient $R_\mathrm{H}$ was obtained from a linear fit of the magnetic field dependence of Hall resistance $(V_\mathrm{H+}-V_\mathrm{H-})/(I_\mathrm{D+}-I_\mathrm{D-})$, where $V_\mathrm{H+}$ and $V_\mathrm{H-}$ are the voltage between the Hall probes of the Hall bar for the positive and negative currents. The Hall carrier density and Hall mobility were obtained as $p_\mathrm{Hall}=1/(e R_\mathrm{H})$ and $\mu_\mathrm{Hall}=R_\mathrm{H}\sigma$. The true carrier density $p$ and drift mobility $\mu$ are given by $p=\gamma_\mathrm{H}p_\mathrm{Hall}$ and $\mu=\mu_\mathrm{Hall}/\gamma_\mathrm{H}$, where $\gamma_\mathrm{H}$ is the Hall scattering factor, which depends on the scattering mechanism and the degree of warping of the valence bands. The Hall scattering factor has been measured and calculated to be around 0.8 for p-type bulk diamond at 300 K. \cite{Mac182} Since no estimates of $\gamma_\mathrm{H}$ have been made for the hole gas confined at the diamond surface, we assumed it to be unity. The output characteristics were measured with a two-point configuration using the source-measure unit (Keysight Technologies, B2901A) for biasing the gate, the function generator (Agilent Technologies, 33220A) for biasing the drain, and the current preamplifier (DL Instruments, 1211) for measuring the drain current. The gate leak current was separately measured by applying a gate voltage and measuring the current through the drain and source by using the current preamplifier. For the electrical measurements of the hydrogen-terminated diamond surface in the glove box, prober needles made of Au-based alloy and with a tip diameter of $20$ ${\mu}$m were directly contacted at diagonal positions of the surface of a hydrogen-terminated diamond substrate ($2.6\times2.6\times0.3$ mm IIa-type (111) single crystalline diamond synthesized by CVD, purchased from Element six; nominal nitrogen concentration below 1 ppm and boron concentration below 0.05 ppm).

\subsection*{Sample preparation for TEM.}

An atomically flat diamond surface was formed on a mesa structure on Ib-type (111) single-crystalline diamond substrate (purchased from Sumitomo Electric Industries) using microwave-plasma-assisted CVD with a low CH$_4$/H$_2$ ratio (0.025\%). \cite{Tok08} The microwave power, pressure, and stage temperature were $800$ W, $80$ Torr, and $810-820$\r{}C, respectively. The flow rate of the H$_2$ gas was $1000$ sccm. AFM measurements confirmed terrace widths of $\approx1$ $\mu$m. The diamond surface was exposed to hydrogen plasma, transferred to the glove box using the vacuum suitcase, and laminated with a flake of h-BN in the glove box, in the same manner as the device fabrication. The h-BN/diamond stack was covered with ${\approx}30$ nm-thick carbon using a carbon evaporation coater (JEOL, JEC-560) for protection and preventing charge-up. For protection against milling, $\approx$1 ${\mu}$m-thick carbon was deposited on the h-BN/diamond stack in the shape of a $12{\times}2$ $\mu$m$^2$ strip with the long side parallel to the $[\bar{1} \bar{1} 2]$ direction of diamond using a focused ion beam (FIB) system (JEOL, JIB-4000). The deposition was performed with a Ga$^{+}$ ion beam (10 kV and 50 pA) under phenanthrene gas injection. Then, $\approx$1 ${\mu}$m-thick carbon was further deposited in the shape of a $30{\times}3$ $\mu$m$^2$ strip covering the $12{\times}2$ $\mu$m$^2$ strip with Ga$^{+}$ (30 kV and 200 pA). In the FIB system, trenches were etched around the larger strip with Ga$^{+}$ (30 kV and 30 nA-700 pA) to cut out a cross-sectional slice with a dimension of $\approx$25 $\mu$m $\times$ 2 $\mu$m $\times$ 5 $\mu$m. The slice was picked up and fixed on a copper grid with epoxy. The slice was further thinned with Ga$^{+}$ (30 kV and 700-100 pA) to a thickness of ${\approx}80$ nm. Finally, gentle milling with Ga$^{+}$ (5 kV and 30 pA) was performed to remove surface damage.

\subsection*{TEM imaging and analysis.}

TEM imaging was carried out using a Cs-corrected TEM (JEOL, JEM-ARM200F with Cold FE-gun) operated at $200$ kV. The $[\bar{1} 1 0]$ crystallographic direction of diamond was adjusted to the electron beam direction. A Fourier transform was performed using DigitalMicrograph(R) software (Gatan Inc.).

\subsection*{Acknowledgements}
We thank H. Osato, E. Watanabe, D. Tsuya, Y. Nishimiya, and F. Uesugi for their technical support. We also thank J. Inoue for helpful discussions, and T. Ando, S. Koizumi, T. Teraji, Y. Wakayama, and T. Nakayama for their support. This study was financially supported by JSPS KAKENHI (Grant Nos. JP19J12696, JP19H05790, JP20H00354, JP19H02605, and JP25287093), the Elemental Strategy Initiative (Grant No. JPMXP0112101001) and the Nanotechnology Platform Project of MEXT, Japan.

\newpage

\vspace{25truecm}

\renewcommand{\theequation}{S\arabic{equation}}
\setcounter{equation}{0}

\textbf{Supplementary Information for ``High-mobility p-channel wide bandgap transistors based on $h$-BN/diamond heterostructures''}

\subsection{Hydrogen-terminated diamond transistors and the transfer doping concept}

Hydrogen-terminated diamond has been widely used to make diamond field-effect transistors (FETs) because of its unique surface conductivity. \cite{Gei18} The hydrogen-terminated surface exposed to air exhibits p-type surface conductivity with a typical sheet resistance of 10-100 k$\Omega$ even when the diamond is not doped intentionally. Since the carrier activation efficiency for the primary dopant, boron, in diamond is low due to its large ionization energy, 0.37 eV, the high conductivity of hydrogen-terminated diamond is attractive for making diamond devices. A widely accepted model for explaining the surface conductivity is based on the transfer doping mechanism. \cite{Mai00} Because of the shallow valence bands of hydrogen-terminated diamond, electrons in the valence bands are transferred to the redox level (2H$_{3}$O$^{+}$ + 2e$^{-}$ $\rightleftharpoons$ H$_{2}$ + 2H$_{2}$O) of a water layer adsorbing on the diamond, which forms a two-dimensional hole gas at the diamond surface. According to this model, the positive charge of the holes, with a density of $10^{12}-10^{13}$ cm$^{-2}$, is balanced by the negative charge of anions, such as HCO$_3^{-}$, in the water layer.

This transfer doping concept has been extended from atmospheric adsorbents to other electron acceptor materials \cite{Str04} and is now used to make diamond FETs. Oxygen point defects and aluminum vacancies in Al$_2$O$_3$ act as electron acceptors in hydrogen-terminated diamond FETs with an Al$_2$O$_3$ gate insulator. \cite{Kaw17} Exposing the diamond to NO$_2$ gas before the deposition of the gate insulator increases the hole density, whereby a high drain current density 1.3 A mm$^{-1}$ has been achieved for a gate length of 0.4 ${\mu}$m. \cite{Kas17} NO$_3^{-}$ was suggested to be the counter charge for holes in this case. \cite{Gei18} High work-function oxides such as V$_2$O$_5$ and an acidic mixed oxide Al$_2$O$_3$-SiO$_2$ have also been used as electron acceptor materials on the diamond surface. \cite{Ver162,Ren17,Yin18,Gei20} Thus, a common strategy for creating holes in hydrogen-terminated diamond has been to introduce empty electron states near the diamond surface that can capture electrons from the diamond valence bands.

These ordinary methods of ``activating'' hydrogen-terminated diamond are accompanied by negative charges near the diamond surface, and they actually limit FET performance. Although the negative charges can create a high density of holes, they also act as sources of carrier scattering and reduce the carrier mobility. \cite{Li18,Sas20} The negative charges also cause a positive shift in the threshold voltage, leading to normally-on operation, \cite{Kas17,Ver162,Ren17,Yin18,Gei20} which is undesirable from the viewpoint of fail-safe operation and driving circuit simplicity. Normally-off diamond FETs have been realized through UV ozone treatment, \cite{Kit17,JFZha20} positive charge accumulation in the gate insulator, \cite{Liu17} and nitrogen implantation in diamond. \cite{Oi19} However, these methods rely on the introduction of defects, and therefore, they degrade transport properties such as the carrier mobility and maximum drain current. Recently, normally-off diamond FETs were fabricated by using other methods, such as through the use of an OH- or Si-terminated surface \cite{Mat19,Fei20} or a metal-insulator-metal-semiconductor structure. \cite{Lia19,MZha20} Some of the device characteristics were improved in these diamond FETs, but the mobility and maximum current were still limited.

The use of $h$-BN and an air-free process in this study substantially reduced the density of surface acceptors and allowed us to make high-hole-mobility diamond transistors with both excellent on-state characteristics and normally-off operation.

\subsection{Comparison of our diamond FETs and prior work}

Figures 5a and 5b in the main text plot the data in the following literature.

Figure 5a: Refs. \cite{Ren17,Yin18,JFZha20,Liu17,Ren18,MZha20,Hok99,Yun99,Ume00,Ume01,Hir10,Hir12,Liu13,Var14,Liu14,Zha16,Liu162,Zha17,Ina19,Wan20,YFWan20,Su20,Sah20} for diamond FETs (brown dots), Ref. \cite{Sas18} for our previous diamond FETs (orange dots), and Refs. \cite{Reu14,Nak10,Cha19,Kri19,Zhe20,Raj20,Kri20} for p-channel GaN-based heterostructures (green dots). 

Figure 5b: Refs. \cite{Ren17,Yin18,Kit17,JFZha20,Oi19,Wan18,Ren18,Fei20,MZha20, Liu13,Var14,Zha16,Liu162,Ina19,Wan20,YFWan20,Su20,Ren172,Lia19,Liu16,Mat16,Sya17,Ren182,Yin182,Mac18,Liu19,YFWan19,YFWan192,YFWan193,Hus20,Ren20,Ren202,Abb20,Lee20,Liu21} for diamond FETs (brown dots), Ref. \cite{Sas18} for our previous diamond FETs (orange dots), Refs. \cite{Reu14,Kri19,Zhe20,Raj20,Kri20,KZha16,Chu16,Bad18,Cho19,Sha02,Li13,Nak18,Cho20} for p-channel GaN FETs (green dots), and Refs. \cite{Con12,Oka12} for p-channel SiC FETs (grey dots).

\subsection{Scattering mechanisms and mobility}

Here, we examine the scattering mechanism that limits the mobility of our FETs. The details of the modelling are described in our previous paper. \cite{Sas20} We calculate the temperature and carrier density dependence of mobility in consideration of five scattering mechanisms: interface charges (interface ionized impurities or interface trapped charges), background ionized impurities, acoustic phonons, nonpolar optical phonons, and surface roughness. The calculations are performed for the first subbands of heavy, light, and split-off holes because the sum of these subbands reaches ${\approx98}\%$ of the total carrier density. Mixing and warping of the valence bands are not considered. We assumed that the hole gas was degenerate in Ref. \cite{Sas20}, whereas here we calculate the averaged relaxation time by taking the Fermi-Dirac distribution into account. Another improvement is that we distinguish the dielectric constant of diamond and that of $h$-BN, whereas we assumed a uniform dielectric medium in Ref. \cite{Sas20}. We also take into account dielectric anisotropy in $h$-BN (Fig. S11a). 

The scattering rate due to charges at a distance $d$ above the diamond surface (Fig. S11a) is given by \cite{And82,Dav98,Mel01}
\begin{eqnarray}
 \frac{1}{\tau _\mathrm{ic}^{i}} = n_\mathrm{ic}\frac{m_{//}^{i}}{2\pi \hbar ^{3}k^{3}}\left(\frac{e^{2}}{2\epsilon _{0} \overline{\epsilon}}\right)^{2}\int_{0}^{2k}[F^{i}(q)]^{2}\frac{1}{[q+q_\mathrm{s}^{i}G^{i}(q)]^{2}}\frac{q^{2}dq}{\sqrt{1-(q/2k)^{2}}},
 \label{eq:si}
\end{eqnarray}
\begin{eqnarray}
 F^{i}(q) = \int_{0}^{\infty}dz\left|\Psi^{i}(z)\right|^{2}\exp \left[-q\left(z+d/\gamma\right)\right],
 \label{eq:FF1}
\end{eqnarray}
\begin{eqnarray}
 G^{i}(q) = \int_{0}^{\infty}dz\int_{0}^{\infty}dz'\left|\Psi^{i}(z) \right|^{2}\left|\Psi^{i}(z') \right|^{2}\left[\exp\left(-q\left|z-z'\right|\right)+\frac{\epsilon_{\mathrm{S}}-\epsilon_{\mathrm{I}}}{\epsilon_{\mathrm{S}}+\epsilon_{\mathrm{I}}}\exp\left[-q\left(z+z'\right)\right]\right], \hspace{5mm}
\label{eq:FormFactor}
\end{eqnarray}
\begin{eqnarray}
q_\mathrm{s}^{i} = \frac{m_{//}^{i}e^{2}}{2\pi \epsilon _{0}\epsilon _\mathrm{S}\hbar ^{2}}\frac{1}{1+\exp[(E_{0}^{i}-E_\mathrm{F})/(k_BT)]},
\end{eqnarray}
\begin{eqnarray}
\gamma=\sqrt{\epsilon_{\mathrm{I}\perp}/\epsilon_{\mathrm{I}//}},
\end{eqnarray}
\begin{eqnarray}
\epsilon_{\mathrm{I}}=\sqrt{\epsilon_{\mathrm{I}\perp}\epsilon_{\mathrm{I}//}},
\end{eqnarray}
\begin{eqnarray}
\overline{\epsilon}=\frac{\epsilon_\mathrm{S}+\epsilon_\mathrm{I}}{2}.
\end{eqnarray}
Here, $e$ is the elementary charge; $\hbar$ is the reduced Planck constant; $k_\mathrm{B}$ is the Boltzmann constant; $T$ is the absolute temperature; $k$ is the wave number; $E_\mathrm{F}$ is the Fermi level; $\epsilon_0$ is the vacuum permittivity; $\epsilon_{\mathrm{S}}$ is the dielectric constant of diamond; $\epsilon_{\mathrm{I}\perp}$ is the dielectric constant of $h$-BN along the direction perpendicular to the layers; and $\epsilon_{\mathrm{I}//}$ is the dielectric constant of $h$-BN along the direction parallel to the layers. We used the following values: $\epsilon_{\mathrm{S}}=5.7$, \cite{Bha48} $\epsilon_{\mathrm{I}\perp}=3$, and $\epsilon_{\mathrm{I}//}=7$. \cite{Ohb01} $\Psi^{i}(z)$ is the wave function of the first subband of heavy (HH), light (LH), and split-off (SO) holes ($i=$ HH, LH, SO); $m_{//}^{i}$ is the effective mass parallel to the diamond surface ($i=$ HH, LH, SO); $n_\mathrm{ic}$ is the sheet density of charges at a distance $d$ above the diamond surface; $q_{s}^{i}$ is the screening wave vector; and $F^{i}(q)$ and $G^{i}(q)$ are form factors. $E_{0}^{i}$ is the minimum energy of holes in the first subband. (The hole energy is taken as positive here. The total energy is $E^{i}+E_{0}^{i} = \hbar^{2}k^{2}/(2m_{//}^{i})+E_{0}^{i}$.) We assume $d=0$ in the following calculation as ionized impurities on the diamond surface or interface trapped charges are the most probable scattering source.

The scattering rate due to background ionized impurities is given by \cite{Dav98}
\begin{eqnarray}
 \frac{1}{\tau _{\rm bc}^{i}} = N_\mathrm{bc}\frac{m_{//}^{i}}{4\pi \hbar ^{3}k^{3}}\left(\frac{e^{2}}{2\epsilon _{0} \epsilon _{\mathrm{S}}}\right)^{2}\int_{0}^{2k}\frac{1}{(q+q_\mathrm{s}^{i})^{2}}\frac{qdq}{\sqrt{1-(q/2k)^{2}}}.
 \label{eq:bi}
\end{eqnarray}
Here, $N_\mathrm{bc}=N_\mathrm{D}^{+}+N_\mathrm{A}^{-}$ is the density of background ionized impurities. Nitrogen (donor) concentration, $N_\mathrm{D}$, and boron (acceptor) concentration, $N_\mathrm{A}$, as estimated from a secondary ion mass spectrometry measurement, are $500$ and $7$ ppb ($8\times10^{16}$ and $1\times10^{15}$ cm$^{-3}$). We assume $N_\mathrm{bc} = 500$ ppb because the donors should be fully ionized near the surface.

The scattering rate due to acoustic phonons is given by \cite{Dav98,Ham17}
\begin{eqnarray}
 \frac{1}{\tau _{\rm ap}^{i}} = \frac{m_{//}^{i}k_\mathrm{B}TD_\mathrm{ap}^{2}}{\rho u_{l}^{2}\hbar ^{3}}\int_{-\infty}^{\infty}\left|\Psi^{i}(z)\right|^{2}\left|\Psi^{i}(z)\right|^{2}dz.
 \label{eq:ac}
\end{eqnarray}
Here, $D_\mathrm{ap}$ is the acoustic deformation potential; $\rho$ is the crystal mass density; and $u_{l}$ is the velocity of longitudinal acoustic phonons. $\rho$ and $u_{l}$ of diamond are 3515 kgm$^{-3}$ and 17536 ms$^{-1}$, respectively. \cite{Per10}

The scattering rate due to nonpolar optical phonons is given by \cite{Ham17}
\begin{eqnarray}
 \frac{1}{\tau _{\rm op}^{i}} = \frac{m_{//}^{i}D_{\rm op}^{2}}{2\rho \hbar ^{2}\omega_{0}}\int_{-\infty}^{\infty}\left|\Psi^{i}(z)\right|^{2}\left|\Psi^{i}(z)\right|^{2}dz \nonumber
\end{eqnarray}
\begin{eqnarray}
\times\left\{N(\omega_{0})\Theta\left[E^{i}(k)+\hbar\omega_{0}\right]+\left(N(\omega_{0})+1\right)\Theta\left[E^{i}(k)-\hbar\omega_{0}\right]\right\},
\end{eqnarray}
\begin{eqnarray}
N(\omega_{0})=\left\{\exp\left[\frac{\hbar \omega_{0}}{k_{B}T}\right]-1\right\}^{-1},
\end{eqnarray}
\begin{eqnarray}
E^{i}(k)=\frac{\hbar^{2}k^{2}}{2m_{//}^{i}}.
\label{eq:op}
\end{eqnarray}
Here, $\Theta\left[x\right]$ is a step function. $D_\mathrm{op}$ is the deformation potential for nonpolar optical phonons, and $\hbar\omega_0$ is the optical phonon energy. $\hbar\omega_0$ of diamond is 165 meV. \cite{Per10}

The scattering rate due to surface roughness is given by \cite{Zan04}
\begin{eqnarray}
 \frac{1}{\tau _\mathrm{sr}^{i}} = \frac{\Delta ^{2}\Lambda^{2}e^{4}m_{//}^{i}}{(\epsilon _{0}\overline{\epsilon})^{2}\hbar ^{3}}(p_\mathrm{2D}+n_\mathrm{depl})^{2}\int_{0}^{1}\frac{u^{4}\exp [-k^{2}\Lambda^{2}u^{2}]}{[u+G^{i}(2ku)q_\mathrm{s}^{i}/(2k)]^{2}\sqrt{1-u^{2}}}du.
 \label{eq:sr}
\end{eqnarray}
The surface roughness is characterized by the average roughness ($\Delta$) and correlation length ($\Lambda$). $p_\mathrm{2D}$ is the total sheet carrier density and $n_\mathrm{depl}$ is the sheet density of the fixed charge in the depletion layer. ($n_\mathrm{depl} = N_\mathrm{depl}z_\mathrm{depl}$, where $N_\mathrm{depl}$ is the volume density of the fixed charge in the depletion layer and $z_\mathrm{depl}$ is the depletion-layer thickness). From the nitrogen and boron concentrations described above, we use $N_\mathrm{depl} = 500$ ppb, for which $z_\mathrm{depl}$ and $n_\mathrm{depl}$ are calculated to be $170$ nm and $1.5\times10^{12}$ cm$^{-2}$, respectively.

The total scattering rate is calculated using the Mathiessen rule,
\begin{eqnarray}
 \frac{1}{\tau^{i}} = 
\frac{1}{\tau _\mathrm{ic}^{i}} + \frac{1}{\tau _\mathrm{bc}^{i}} + \frac{1}{\tau _\mathrm{ap}^{i}} + \frac{1}{\tau _\mathrm{op}^{i}} + \frac{1}{\tau _\mathrm{sr}^{i}}.
 \label{eq:tauall}
\end{eqnarray}
The mobility for $i=$ HH, LH, and SO is obtained from $\mu^{i} = e\left<\tau^{i}\right>/m_{//}^{i}$, where $\left<\tau^{i}\right>$ is the energy averaged relaxation time,
\begin{eqnarray}
  \left<\tau^{i}\right> = \frac{\displaystyle \int_{0}^{\infty}\tau^{i}E^{i}\frac{df(E^{i}+E_{0}^{i})}{dE^{i}}dE^{i}}{\displaystyle \int_{0}^{\infty}E^{i}\frac{df(E^{i}+E_{0}^{i})}{dE^{i}}dE^{i}}.
\label{eq:tauallave}
\end{eqnarray}
Here, $f(E)$ is the Fermi-Dirac distribution function.

Finally, the mobility $\mu$ and carrier density $p$ of the FET are calculated using the formula for the multi-carrier Hall effect,
\begin{eqnarray}
 \mu = \frac{p^{\rm HH}(\mu^{\rm HH})^{2}+p^{\rm LH}(\mu^{\rm LH})^{2}+p^{\rm SO}(\mu^{\rm SO})^{2}}{p^{\rm HH}\mu^{\rm HH}+p^{\rm LH}\mu^{\rm LH}+p^{\rm SO}\mu^{\rm SO}},
 \label{eq:muHall}
\end{eqnarray}
\begin{eqnarray}
 p = \frac{(p^{\rm HH}\mu^{\rm HH}+p^{\rm LH}\mu^{\rm LH}+p^{\rm SO}\mu^{\rm SO})^{2}}{p^{\rm HH}(\mu^{\rm HH})^{2}+p^{\rm LH}(\mu^{\rm LH})^{2}+p^{\rm SO}(\mu^{\rm SO})^{2}}.
 \label{eq:nHall}
\end{eqnarray}

Figure S12a shows the temperature dependence of mobility of device C1 at $V_\mathrm{GS} = - 8$ V. A reasonable match between the measured and calculated mobilities is obtained for the following parameters: $n_\mathrm{ic} = 6\times10^{11}$ cm$^{-2}$, $D_\mathrm{ap} = 8$ eV, $D_\mathrm{op} = 1.5\times10^{10}$ eV/cm, $\Delta = 0.18$ nm, and $\Lambda = 2$ nm. The value of $n_\mathrm{ic}$ is lower than the one that was estimated for air-exposed samples in our previous study. \cite{Sas20} The values of $\Delta$, $\Lambda$, $D_\mathrm{ap}$, and $D_\mathrm{op}$ are comparable to previously reported ones. \cite{Li18,Sas20,Dal20,Pet20} The figure indicates that the mobility is limited by the interface charges and surface roughness at low temperatures, whereas the acoustic and nonpolar optical phonons also contribute to the carrier scattering and mobility reduction at room temperature. Figure S12b shows the carrier-density dependence of mobility of devices C1 and C2 at 300 K. It also shows the carrier-density dependence of mobility calculated with the same parameter sets used for the calculation of the temperature dependence of mobility in Fig. S12a. The increase in mobility with increasing carrier density is attributed to the carrier scattering due to the remaining interface charges.

\subsection{Gate voltage dependence of carrier density}

In this section, we examine the gate voltage dependence of carrier density and the origin of the normally-off behavior (i.e. negative threshold voltage). The relation between the hole density and gate voltage $V_\mathrm{GS}$ is expressed as
\begin{eqnarray}
V_\mathrm{GS} = -\frac{e(p_\mathrm{2D} + n_\mathrm{depl} - n_\mathrm{ic}) t_\mathrm{hBN}}{\epsilon_\mathrm{hBN}} + \psi_\mathrm{s} + \phi_\mathrm{ms}.
\end{eqnarray}
Here, $p_\mathrm{2D}$ is the total sheet carrier density. $n_\mathrm{depl}$ is the sheet density of the fixed charge in the depletion layer. $n_\mathrm{ic}$ is the sheet density of negative charges at the diamond surface. $t_\mathrm{hBN}$ and $\epsilon_\mathrm{hBN}$ are the thickness and dielectric constant of the h-BN gate insulator. $\psi_\mathrm{s}$ ($\textless0$) is the surface potential (relative to deep inside the diamond). $e\phi_\mathrm{ms}= e\phi_\mathrm{m} - e\phi_\mathrm{s}$ is the difference between the work function ($e\phi_\mathrm{m}$) of the graphite gate and that ($e\phi_\mathrm{s}$) of hydrogen-terminated diamond. $e\phi_\mathrm{s} \approx 0.3$ eV, because the electron affinity of hydrogen-terminated diamond is $-1.3$ eV \cite{Cui98} and the Fermi level in the bulk diamond with $N_\mathrm{D}$ (nitrogen) = 500 ppb and $N_\mathrm{A}$ = 7 ppb is $-1.6$ eV below the conduction band minimum. (The nitrogen donor ionization energy in diamond is $1.7$ eV. \cite{Far69}) $e\phi_\mathrm{m}$ for graphite is 4.7 eV. \cite{Rut20} Therefore, $e\phi_\mathrm{ms} \approx 4.4$ eV. For a given $p_\mathrm{2D}$, $n_\mathrm{depl}$ and $\psi_\mathrm{s}$ can be evaluated by self-consistently solving the Schr\"{o}dinger and Poisson equations. $n_\mathrm{ic}$ can be determined by comparing the experimental and calculated mobilities, as described above. 

Figure S12c shows $V_\mathrm{GS}-p_\mathrm{2D}$ curves calculated with two different parameter sets. The calculated curves are in reasonable agreement with the experimental one. However, a slightly lower value ($4\times10^{11}$ cm$^{-2}$) of $n_\mathrm{ic}$ than the one ($6\times10^{11}$ cm$^{-2}$) that explains the temperature dependence of mobility (Fig. S12a) provides a better fit to the measured $V_\mathrm{GS}-p_\mathrm{2D}$ curve. This can be explained if both interface negative charges due to remaining adsorbates and positive charges of donor-type interface states contribute to the reduction in mobility and the shift in threshold voltage. The interface charge density $n_\mathrm{ic}$ in the formula (Eq. S1) for the scattering rate is the sum of the negative and positive charge densities ($n_\mathrm{ic} = n_\mathrm{ic}^{-} + n_\mathrm{ic}^{+}$), whereas $n_\mathrm{ic}$ in the formula (Eq. S18) for the gate voltage dependence of the carrier density is the net density of negative charges ($n_\mathrm{ic} = n_\mathrm{ic}^{-} - n_\mathrm{ic}^{+}$). Therefore, a negative charge density of $5\times10^{11}$ cm$^{-2}$ and a positive charge density of $1\times10^{11}$ cm$^{-2}$ can explain both the temperature dependence of mobility and the gate voltage dependence of carrier density in a consistent manner. This indicates that negative charges due to remaining adsorbates mainly cause the reduction in mobility, but that ionized donor-type interface states also contribute to it.

Energy-band diagrams of the graphite/h-BN/hydrogen-terminated diamond heterostructure were calculated using reported values of the electron affinity, ${\approx}1$ eV, \cite{Liu15} and bandgap, $6.0$ eV, \cite{Wat04} of h-BN, dielectric constant of diamond, 5.7, \cite{Bha48} and dielectric constant of $h$-BN, $3.3$, \cite{You12,Ste20}, in addition to the parameters described above. Figure S13 shows the energy-band diagrams for two different densities of interface negative charges, $n_\mathrm{ic} = 4\times10^{11}$ and $5\times10^{12}$ cm$^{-2}$, indicating normally-off and normally-on operations, respectively. For $n_\mathrm{ic} = 5\times10^{12}$ cm$^{-2}$, an inversion layer of holes ($\approx3.7\times10^{12}$ cm$^{-2}$) forms even when $V_\mathrm{GS} = 0$ V. This normally-on operation for large $n_\mathrm{ic}$ corresponds to that of ordinary hydrogen-terminated diamond FETs based on surface transfer doping. Figures 1e and 1f in the main text show energy-band diagrams for $n_\mathrm{ic}=0$. Even in this case, an inversion layer of holes is generated by applying a gate voltage, indicating that surface transfer doping is unnecessary for operating hydrogen-terminated diamond FETs.

\subsection{Modelling of output characteristics}

The output characteristics are modelled by a charge-sheet model within the gradual-channel approximation. \cite{Bre78,Sze07} The drain current is given by
\begin{eqnarray}
\frac{I_\mathrm{D}}{W_\mathrm{G}} = -\frac{\mu^{*} \epsilon_\mathrm{hBN}}{L_\mathrm{G} t_\mathrm{hBN}}\left(V_\mathrm{GS}- \phi_\mathrm{ms}-\frac{e n_\mathrm{ic} t_\mathrm{hBN}}{\epsilon_\mathrm{hBN}}-\psi_\mathrm{s0}-\frac{V_\mathrm{DS}}{2}\right)V_\mathrm{DS}\nonumber
\end{eqnarray}
\begin{eqnarray}
+\frac{2\mu^{*}}{3L_\mathrm{G}}\left(2e\epsilon_\mathrm{S}N_\mathrm{D}\right)^{1/2}\left[\left|V_\mathrm{DS}+\psi_\mathrm{s0}\right|^{3/2}-\left|\psi_\mathrm{s0}\right|^{3/2}\right],
\end{eqnarray}
where $\psi_\mathrm{s0}$ is the surface potential at the source end of the channel. $\mu^{*}$ is an effective mobility that is assumed to be constant over the channel. Note that $I_\mathrm{D}$, $V_\mathrm{GS}$, $V_\mathrm{DS}$, and $\psi_\mathrm{s0}$ are negative, while $\phi_\mathrm{ms}$ is positive. The absolute values of drain voltage $V_\mathrm{DS}$ and gate voltage $V_\mathrm{GS}$ are smaller than those of the applied drain voltage, $V_\mathrm{DS}^a$, and applied gate voltage, $V_\mathrm{GS}^a$, respectively, because of the contact resistances ($R_\mathrm{cS}$ and $R_\mathrm{cD}$) of the source and drain electrodes. \cite{SKim12} (Fig. S11b)
\begin{eqnarray}
V_\mathrm{DS} = V_\mathrm{DS}^a - (R_\mathrm{cS}+ R_\mathrm{cD}) I_\mathrm{D},
\end{eqnarray}
\begin{eqnarray}
V_\mathrm{GS} = V_\mathrm{GS}^a - R_\mathrm{cS}I_\mathrm{D}.
\end{eqnarray}
The effect of the contact resistance on the gate voltage is more significant for larger gate voltages, i.e. for larger drain currents. For example, a contact resistance of $5$ ${\Omega}$mm and a drain current of $200$ mA/mm lead to a gate voltage reduction of $1$ V. The measured output characteristics are well explained by the $I_\mathrm{D}-V_\mathrm{DS}^a$ curves calculated for $V_\mathrm{DS}^a=0-10$ V using the above equations (Fig. 3e in the main text). For the calculation, we used the contact resistance $R_\mathrm{cS}+R_\mathrm{cD}$ estimated from two- and four-point resistances of the device (Fig. S10) and assumed $R_\mathrm{cS}=0.7(R_\mathrm{cS}+R_\mathrm{cD})$. The following parameters were also used: $\mu^{*}=680$ cm$^{2}$V$^{-1}$s$^{-1}$, $L_\mathrm{G} = 8.09$ $\mu$m, $\epsilon_\mathrm{hBN}/\epsilon_0=3.3$, $t_\mathrm{hBN}=24$ nm, $\epsilon_\mathrm{S}/\epsilon_0=5.7$, $e\phi_\mathrm{ms} = 4.4$ eV, $N_\mathrm{D}=9\times10^{16}$ cm$^{-3}$, $\psi_\mathrm{s0} = -4.1$ V, and $n_\mathrm{ic} = 4\times10^{11}$ cm$^{-2}$. These values are consistent with the experiments and the modelling of the mobility and threshold voltage described above.

\subsection{Effective and field-effect mobilities}

The field-effect mobility $\mu_\mathrm{FE}$ ($\mu_{\mathrm{FE}}= \frac{t_{\mathrm{hBN}}}{\epsilon _{\mathrm{hBN}}}\left|\frac{\partial \sigma}{\partial V_{\mathrm{GS}}}\right|$) is generally related to the effective mobility $\mu_\mathrm{eff}$ ($\mu_{\mathrm{eff}}= \frac{t_{\mathrm{hBN}}}{\epsilon _{\mathrm{hBN}}}\frac{\sigma}{\left|V_{\mathrm{GS}} -V_{\mathrm{th}}\right|}$) by \cite{Kan89}
\begin{eqnarray}
\mu_{\mathrm{FE}}= {\mu}_{\mathrm{eff}} + (V_{\mathrm{GS}}-V_{\mathrm{th}})\frac{\partial {\mu}_{\mathrm{eff}}}{\partial V_{\mathrm{GS}}},
\end{eqnarray}
for both n-channel FETs ($V_{\mathrm{GS}}-V_{\mathrm{th}}{\textgreater}0$) and p-channel FETs ($V_{\mathrm{GS}}-V_{\mathrm{th}}{\textless}0$). $\mu_\mathrm{FE}$ of our FETs is larger than $\mu_\mathrm{eff}$, as shown in Fig. 3a. This is because the sign of $(V_{\mathrm{GS}}-V_{\mathrm{th}})\frac{\partial {\mu}_{\mathrm{eff}}}{\partial V_{\mathrm{GS}}}$ is positive ($V_{\mathrm{GS}}-V_{\mathrm{th}}{\textless}0$ and $\frac{\partial {\mu}_{\mathrm{eff}}}{\partial V_{\mathrm{GS}}}{\textless}0$). The negative sign of $\frac{\partial {\mu}_{\mathrm{eff}}}{\partial V_{\mathrm{GS}}}$ of our p-channel FETs indicates that the mobility increases with increasing carrier density. This behavior is attributed to the carrier scattering caused by the remaining interface charges, as shown in Supplementary Section C and Fig. S12b. In n-channel Si MOSFETs, the mobility usually decreases with increasing carrier density because phonon and surface roughness scattering are the dominant sources of carrier scattering. Therefore, the sign of $(V_{\mathrm{GS}}-V_{\mathrm{th}})\frac{\partial {\mu}_{\mathrm{eff}}}{\partial V_{\mathrm{GS}}}$ is negative ($V_{\mathrm{GS}}-V_{\mathrm{th}}{\textgreater}0$ and $\frac{\partial {\mu}_{\mathrm{eff}}}{\partial V_{\mathrm{GS}}}{\textless}0$) and $\mu_\mathrm{FE}$ is smaller than $\mu_\mathrm{eff}$, as shown in Ref. \cite{Kan89}.

\newpage

\begin{table}
\begin{tabular}{|c|c|c|c|c|c|c|c|c|c|c|}
\hline
Device & $L_\mathrm{G}$ & $L_\mathrm{G}^{*}$ & $W_\mathrm{G}$ & $L_\mathrm{p}$ & $t_\mathrm{hBN}$ & $V_\mathrm{th}$ & $\mu_\mathrm{Hall}^\mathrm{max}$ & $\rho^\mathrm{min}$ & $L_\mathrm{G}^{*} I_\mathrm{D}^\mathrm{max} / W_\mathrm{G}$ & on/off ratio\\
 & $\mu$m & $\mu$m & $\mu$m & $\mu$m & nm & V & cm$^2$V$^{-1}$s$^{-1}$ & k$\Omega$ & $\mu$m mA mm$^{-1}$ & ($V_\mathrm{DS}=-10$ V)\\
\hline\hline
C1 & 9.10 & 8.09 & 0.89 & 2.80 & 24 & 0.99 & 680 & 1.4 & 1600 & $4.7\times10^{8}$ \\
\hline
C2 & 8.20 & 7.17 & 0.85 & 2.54 & 24 & 0.90 & 620 & 1.6 & 2000 & $1.5\times10^{6}$ \\
\hline
\end{tabular}
\begin{flushleft}
{\footnotesize Table S1: \textbf{Parameters of devices C1 and C2.} $L_\mathrm{G}$, channel length; $L_\mathrm{G}^{*}$, effective channel length; $W_\mathrm{G}$, channel width; $L_\mathrm{p}$, distance between voltage probes; $t_\mathrm{hBN}$, thickness of $h$-BN gate insulator; $V_\mathrm{th}$, threshold voltage ; $\mu_\mathrm{Hall}^\mathrm{max}$, maximum Hall mobility; $\rho^\mathrm{min}$, minimum sheet resistance; $L_\mathrm{G}^{*} I_\mathrm{D}^\mathrm{max} / W_\mathrm{G}$, maximum normalized drain current. $L_\mathrm{G}$, $W_\mathrm{G}$, and $L_\mathrm{p}$ were evaluated from the optical microscope images of the devices (Fig. S1). The effective channel length $L_\mathrm{G}^{*}$ takes into account the spread of the channel width near the source and drain electrodes: $L_\mathrm{G}^{*}=L_\mathrm{G}-L_\mathrm{G}^\mathrm{S}-L_\mathrm{G}^\mathrm{D}+ (L_\mathrm{G}^\mathrm{S})^2W_\mathrm{G}/\int_{0}^{L_\mathrm{G}^\mathrm{S}} W_\mathrm{G}^\mathrm{S}(y)dy+ (L_\mathrm{G}^\mathrm{D})^2W_\mathrm{G}/\int_{0}^{L_\mathrm{G}^\mathrm{D}} W_\mathrm{G}^\mathrm{D}(y)dy$, where $W_\mathrm{G}^\mathrm{S}(y)$ and $W_\mathrm{G}^\mathrm{D}(y)$ are the channel width near the source and drain electrodes, and $L_\mathrm{G}^\mathrm{S}$ and $L_\mathrm{G}^\mathrm{D}$ are the corresponding length for which the channel width is larger than $W_\mathrm{G}$. $t_\mathrm{hBN}$ was evaluated from the atomic force microscope images of the devices (Fig. S2). The electrical characteristics of device C1 are shown in Figs. 3, 4 and 5 and Supplementary Figs. S6, S7, S10, S12 and S13, and those of device C2 are shown in Fig. 5 and Supplementary Figs. S8, S9, S10 and S12. The mobility slightly increased after the measurement of the output characteristics, which led to a slight difference (within 12\%) in mobility between Figs. 3a and 4a and between Figs. S8a and S9b. This behavior might have been caused by relocation of residual surface acceptors under the electric fields. The initial lower mobilities are shown in this table. The lower maximum drain current of C1 than that of C2 despite the higher mobility is attributed to the higher source contact resistance (Supplementary Section E and Fig. S10).}
\end{flushleft}
\end{table}

\begin{table}
\scriptsize
\begin{tabular}{|c|c|c|c|c|c|c|c|c|}
\hline
Ref. & $\mu^\mathrm{max}$ & $p(\mu^\mathrm{max})$ & Mobility extraction & Orientation & Mode & Gate insulator & Gate & Source/Drain \\
 & cm$^2$V$^{-1}$s$^{-1}$ & $10^{12}$ cm$^{-2}$ & & & & & & \\
\hline\hline
\cite{Ren17} & 108 & 5.7 & eff, $C_\mathrm{FET}$ & - & on & MoO$_3$ & Al & Au \\
\hline
\cite{Yin18} & 20.2 & 5.1 & FE, $\epsilon_\mathrm{liter}$ & (001) & on & H$_y$MO$_{3-x}$/HfO$_2$ & Ti/Au & Ti/Au \\
\hline
\cite{JFZha20} & 276 & 0.47 & eff, ${\int}C_\mathrm{FET}$ & (001) & off & Al$_2$O$_3$ & Al & Au \\
\hline
\cite{Liu17} & 35.1 & 13 & eff, $C_\mathrm{MIS}$ & (001) & off & Y$_2$O$_3$ & Ti/Au & Pd/Ti/Au \\
\hline
\cite{Ren18} & 94.2 & 0.33 & eff, ${\int}C_\mathrm{FET}$ & (001) & off & alumina & Al & Au \\
\hline
\cite{MZha20} & 313 & 1.3 & sat, $C_\mathrm{FET}$ & (001) & off &  & Ti/TiO$_x$/Al & Au \\
\hline
\cite{Hok99} & 120 & 2.4 & eff, $\epsilon_\mathrm{liter}$ & (001) & on & SiO$_2$ & Pb & Au \\
\hline
\cite{Yun99} & 400 & 1.3 & eff, - & (001) & on & CaF$_2$ & Al & Au \\
\hline
\cite{Ume00} & 280 & 1.1 & FE, $C_\mathrm{MIM}$ & (001) & on & CaF$_2$ & Cu & Au \\
\hline
\cite{Ume01} & 200 & 0.55 & FE, $C_\mathrm{MIM}$ & (001) & on & CaF$_2$ & Cu & Au \\
\hline
\cite{Hir10} & 143 & 0.93 & - & (001) & - & Al$_2$O$_3$ & Al & Au \\
\hline
\cite{Hir10} & 95 & 2.4 & - & (111) & on & Al$_2$O$_3$ & Al & Au \\
\hline
\cite{Hir12} & 110 & 33 & eff, - & poly & on & NO$_2$/Al$_2$O$_3$ & Al & Au \\
\hline
\cite{Liu13} & 38.7 & 12 & eff, $C_\mathrm{MIS}$ & (001) & off & HfO$_2$ & Pd/Ti/Au & Pd/Ti/Au \\
\hline
\cite{Var14} & 32 & 3.3 & eff, ${\int}C_\mathrm{FET}$ & (001) & off & MoO$_3$/HfO$_2$ & Ti/Au & Ti/Au \\
\hline
\cite{Liu14} & 217.5 & 3.74 & eff, $C_\mathrm{MIS}$ & (001) & on & Al$_2$O$_3$/ZrO$_2$ & Pd/Ti/Au & Pd/Ti/Au \\
\hline
\cite{Zha16} & 88 & 6.1 & eff, $C_\mathrm{MIS}$ & (001) & off & TiO$_x$ & WC & Pd/Ti/Au \\
\hline
\cite{Liu162} & 34.2 & 20 & eff, $\epsilon_\mathrm{liter}$ & (001) & on & Al$_2$O$_3$/ZrO$_2$ & Pd/Ti/Au & Pd/Ti/Au \\
\hline
\cite{Zha17} & 482 & 0.31 & eff, ${\int}C_\mathrm{FET} $ & - & on & MoO$_3$ & Al & Au \\
\hline
\cite{Ina19} & 25 & 1 & sat, $\epsilon_\mathrm{liter}$ & (001) & on & vacuum/SiO$_2$ & n-Si & Ti(TiC)/Pt/Au \\
\hline
\cite{Wan20} & 195 & 3.25 & sat, $C_\mathrm{FET}$ & (001) & off &  & LaB$_6$/Al & Au \\
\hline
\cite{YFWan20} & 145 & 2.1 & sat, $C_\mathrm{FET}$ & (001) & on & Al$_2$O$_3$/LiF & Al & Au \\
\hline
\cite{Su20} & 18.7 & 8.74 & FE, $C_\mathrm{FET}$ & poly & off & Al$_2$O$_3$/HfZrO$_x$ & Al & Au \\
\hline
\cite{Sah20} & 32.87 & 35.1 & eff, $C_\mathrm{MIM}$ & (001) & on & NO$_2$/Al$_2$O$_3$ & Au & Au \\
\hline
\cite{Sas18} & 320 & 6.9 & Hall & (111) & on & h-BN & Ti/Au & Ti(TiC)/Pt \\
\hline
\cite{Sas18} & 315 & 7.2 & Hall & (111) & on & h-BN & Ti/Au & Ti(TiC)/Pt \\
\hline
\cite{Sas18} & 439 & 1.6 & Hall & (111) & on & h-BN & Ti/Au & Ti(TiC)/Pt \\
\hline
This work (C1) & 680 & 6.6 & Hall & (111) & off & h-BN & graphite & Ti(TiC)/Pt \\
\hline
This work (C2) & 620 & 6.4 & Hall & (111) & off & h-BN & graphite & Ti(TiC)/Pt \\
\hline
\end{tabular}
\end{table}
\begin{table}
\begin{flushleft}
{\footnotesize Table S2: \textbf{Summary of room-temperature mobilities and mobility extraction methods for the diamond FETs shown in Fig. 5a.} Ref., reference number; $\mu^\mathrm{max}$ and $p(\mu^\mathrm{max})$, maximum hole mobility and corresponding hole sheet density; Orientation, orientation of diamond; Mode, normally-on or normally-off operation. ``eff", ``FE", ``sat" and ``Hall" in the mobility extraction methods denote effective mobility $\mu_\mathrm{eff}$, field-effect mobility $\mu_\mathrm{FE}$, saturation mobility $\mu_\mathrm{sat}$, and Hall mobility $\mu_\mathrm{Hall}$, respectively: $\mu_\mathrm{eff}= \frac{L_\mathrm{G}}{W _\mathrm{G}}\frac{1}{e p}(\frac{{\partial}V_\mathrm{DS}}{{\partial}I_\mathrm{D}}-R_\mathrm{S}-R_\mathrm{D})^{-1}$. $\mu_\mathrm{FE}= \frac{L_\mathrm{G}}{W _\mathrm{G}}\frac{1}{C _\mathrm{i}}\left|\frac{1}{V_\mathrm{DS}}\frac{{\partial}I_\mathrm{D}}{{\partial}V_\mathrm{GS}}\right|$. $\mu_\mathrm{sat}= \frac{2L_\mathrm{G}}{W _\mathrm{G}}\frac{1}{C _\mathrm{i}}\frac{I_\mathrm{D}^\mathrm{sat}}{(V_\mathrm{GS} -V_\mathrm{th})^2}$. Here, $L_\mathrm{G}$, gate length; $W_\mathrm{G}$, gate width; $C_\mathrm{i}$, gate capacitance; $I_\mathrm{D}$, drain current; $I_\mathrm{D}^\mathrm{sat}$, the absolute value of saturation drain current; $V_\mathrm{DS}$, drain voltage; $V_\mathrm{GS}$, gate voltage; $V_\mathrm{th}$, threshold voltage; $R_\mathrm{S}$ and $R_\mathrm{D}$, source and drain series resistances including contact and access resistances; $e$, the elementary charge; $p$, hole sheet density. $C_\mathrm{i}$ is evaluated either from the capacitance-voltage measurement of the FET (denoted as $C_\mathrm{FET}$), or of a metal-insulator-semiconductor structure fabricated separately ($C_\mathrm{MIS}$), or of a metal-insulator-metal structure fabricated separately ($C_\mathrm{MIM}$), or from the thickness and the literature value of the dielectric constant of the gate insulator ($\epsilon _\mathrm{liter}$). For the effective mobility, $p=\frac{1}{e}C_\mathrm{i}\left|V_\mathrm{GS} -V_\mathrm{th}\right|$ (denoted as  $C_\mathrm{FET}$, $C_\mathrm{MIS}$, $C_\mathrm{MIM}$, or $\epsilon _\mathrm{liter}$ depending on the evaluation method of $C_\mathrm{i}$), or $p=\frac{1}{e}{\int}C_\mathrm{FET}dV_\mathrm{GS}$ (denoted as ${\int}C_\mathrm{FET}$). The stacks of the gate insulator and electrodes are represented in the order of film deposition, i.e., (bottom layer)/(top layer). The FETs in Refs. \cite{MZha20} and \cite{Wan20} are Schottky-gated. ``poly" denotes polycrystalline diamond. Some of $p(\mu^\mathrm{max})$ are not shown in the paper, but calculated using given parameters such as $C_\mathrm{i}$. The items represented by - are not explicitly specified in the paper.}
\end{flushleft}
\end{table}
\newpage

\begin{figure}
\includegraphics[width=8.5truecm]{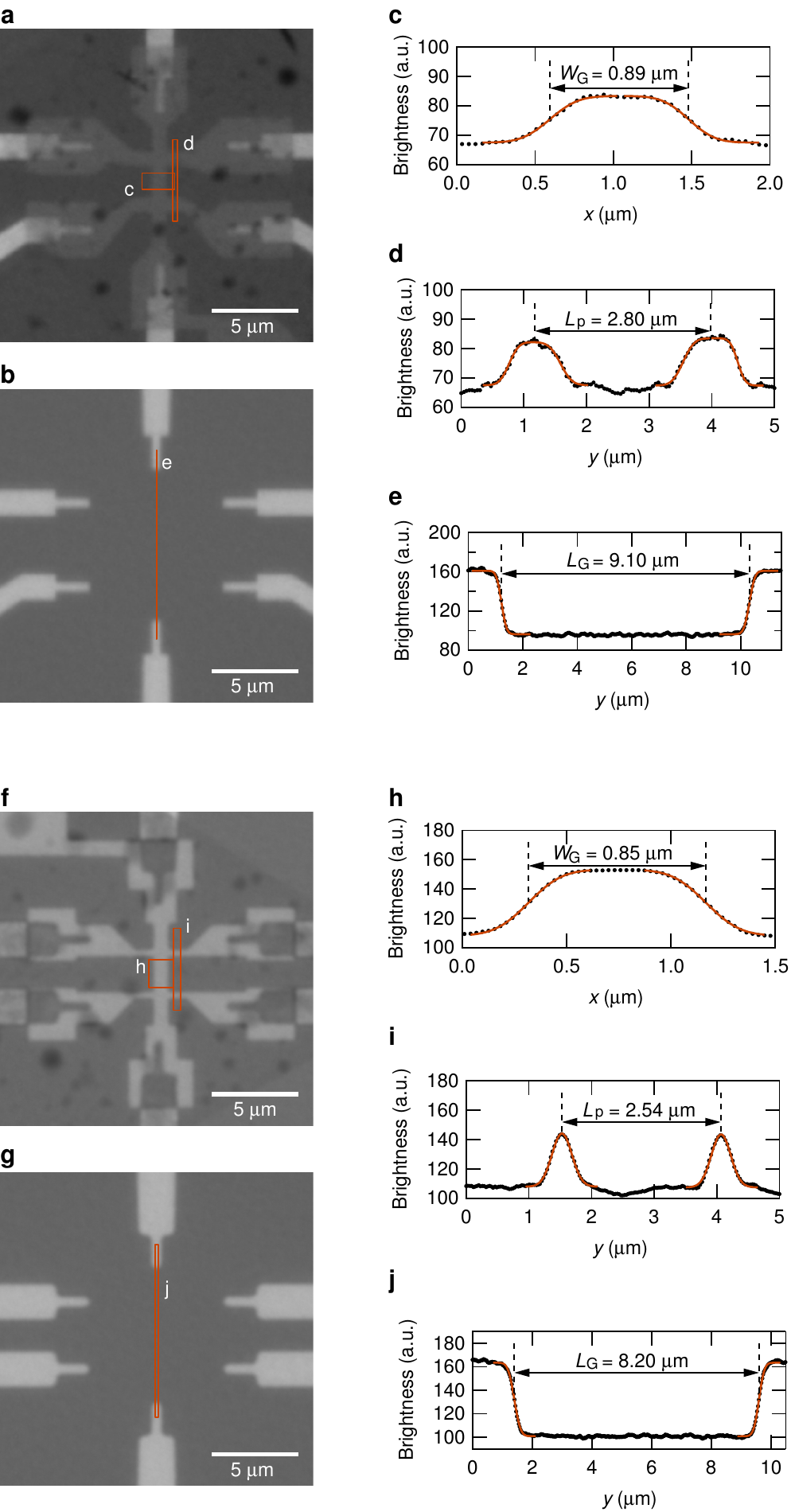}
\begin{flushleft}
{\footnotesize Figure S1: \textbf{Evaluation of geometrical parameters.} \textbf{a,b,f,g}, optical micrograph images of devices C1 (\textbf{a,b}) and C2 (\textbf{f,g}). \textbf{b} and \textbf{g} were taken before the $h$-BN lamination. \textbf{c,h}, image brightness averaged over the $y$ axis of the square labelled c (h) in \textbf{a} (\textbf{f}) is plotted as a function of $x$. The step edge is fitted with the error function, a convolution of gaussian and step functions. The distance between the midpoints of the two error-function fits is taken as the channel width ($W_\mathrm{G}$). \textbf{d,i}, image brightness averaged over the $x$ axis of the square labelled d (i) in \textbf{a} (\textbf{f}) is plotted as a function of $y$. The distance between voltage probes ($L_\mathrm{p}$) is obtained from fits with the error function (\textbf{d}) and gaussian (\textbf{i}). \textbf{e,j}, image brightness averaged over the $x$ axis of the square labelled e (j) in \textbf{b} (\textbf{g}) is plotted as a function of $y$. The distance between the midpoints of the two error-function fits is taken as the channel length ($L_\mathrm{G}$).}
\end{flushleft}
\end{figure}

\begin{figure}
\includegraphics[width=12truecm]{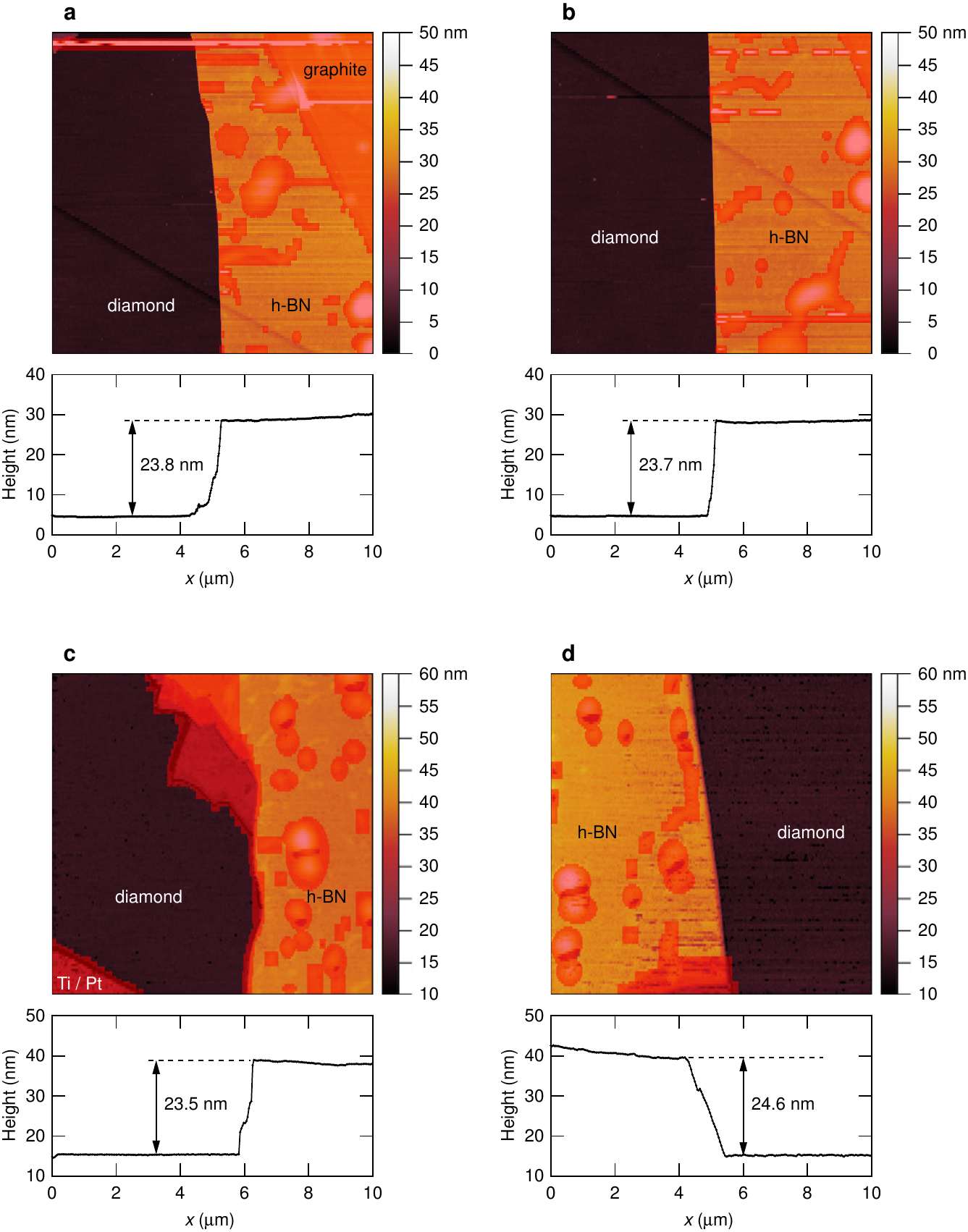}
\begin{flushleft}
{\footnotesize Figure S2: \textbf{Evaluation of the thickness of $h$-BN.} Atomic force microscope (AFM) images of $h$-BN of devices C1 (\textbf{a}, \textbf{b}) and C2 (\textbf{c}, \textbf{d}) before etching. The bottom figures show the height profile along the horizontal direction of the AFM images. The height was averaged over the vertical direction of the images. The areas of the graphite gate, Ti / Pt electrode, and contamination bubbles were masked.}
\end{flushleft}
\end{figure}

\begin{figure}
\includegraphics[width=14truecm]{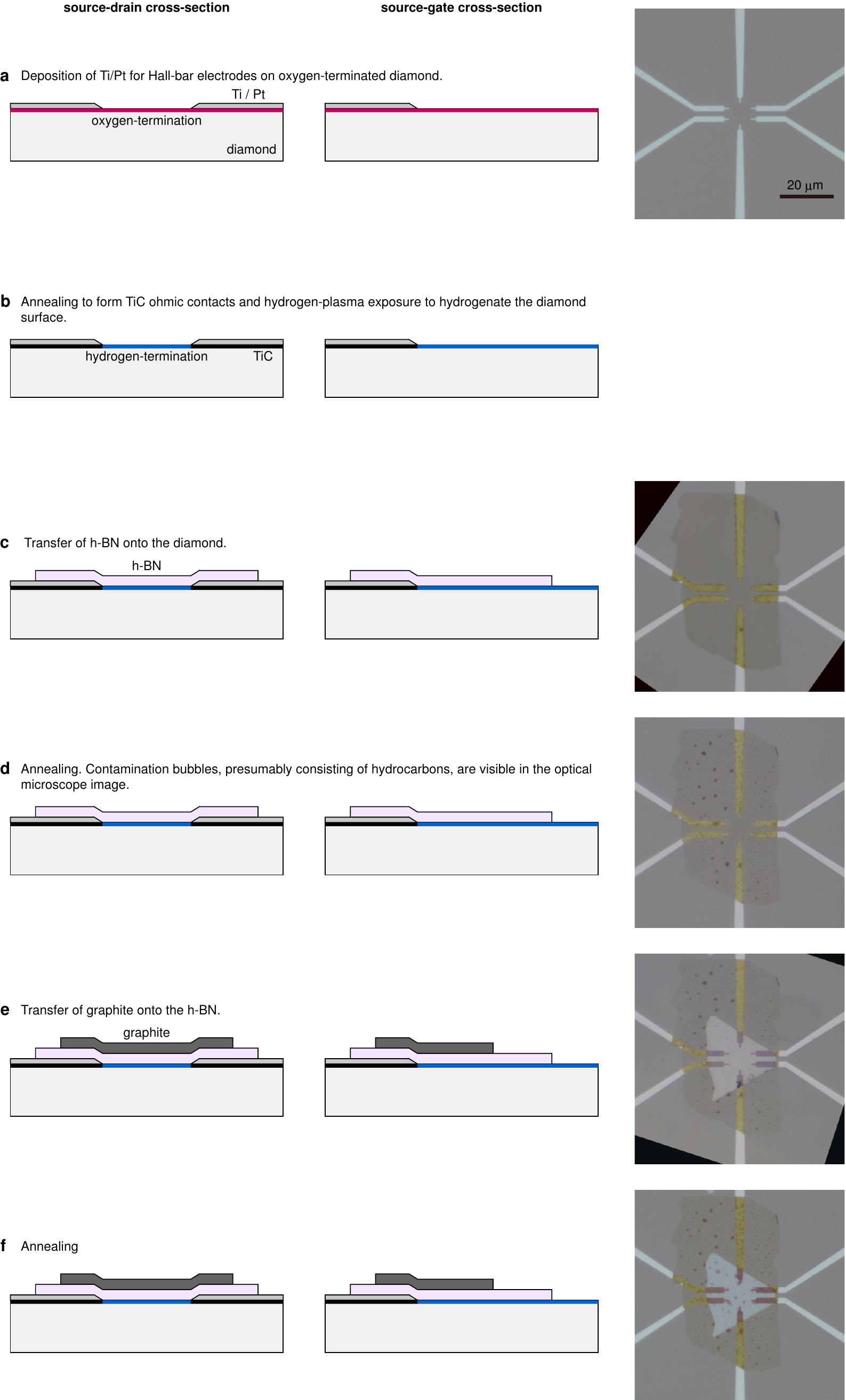}\\
\end{figure}

\begin{figure}
\includegraphics[width=14truecm]{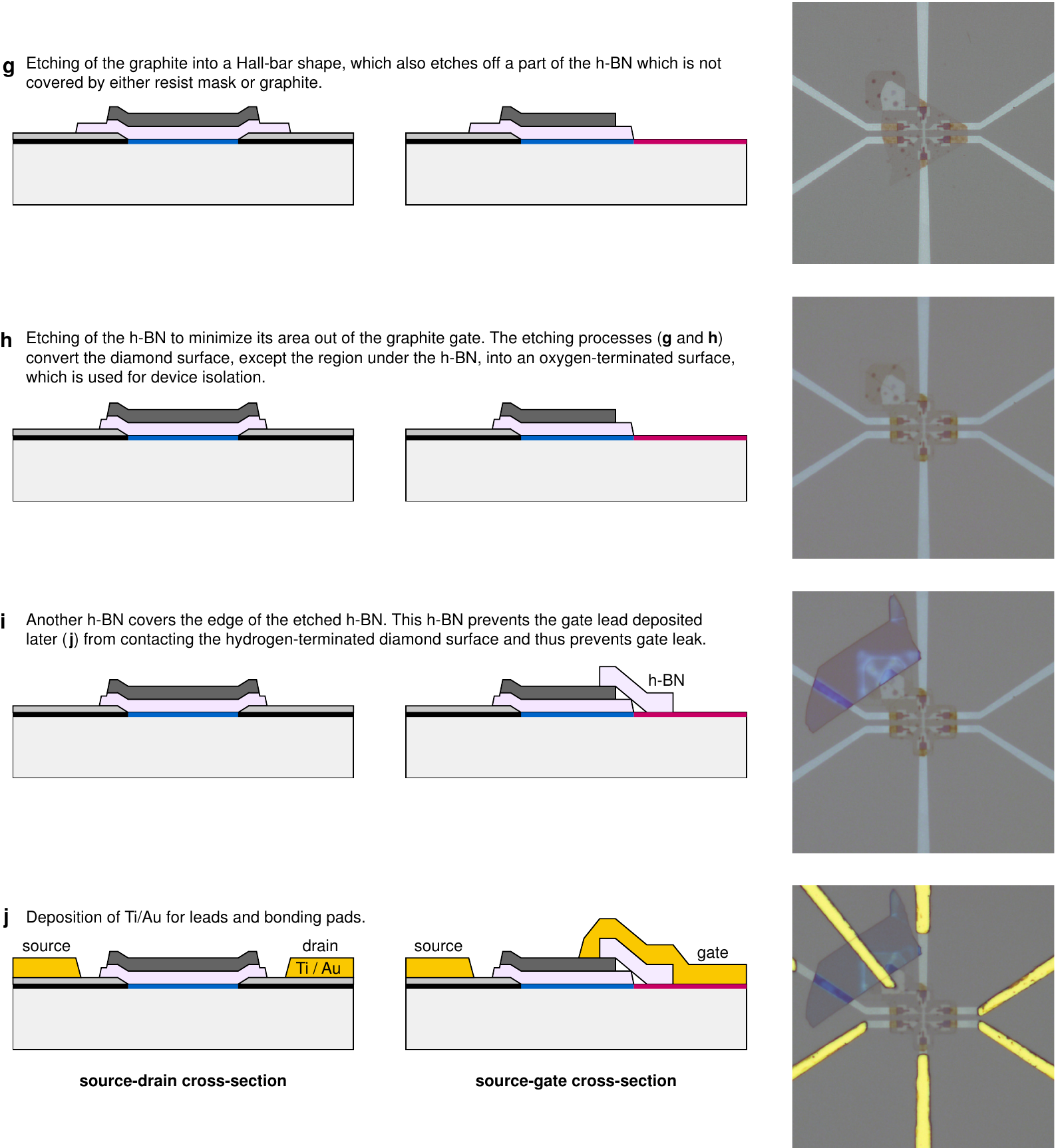}\\
\begin{flushleft}
{\footnotesize Figure S3: \textbf{Schematic diagrams of device fabrication process and optical images taken after each step.}}
\end{flushleft}
\end{figure}

\newpage

\begin{figure}
\includegraphics[width=11.2truecm]{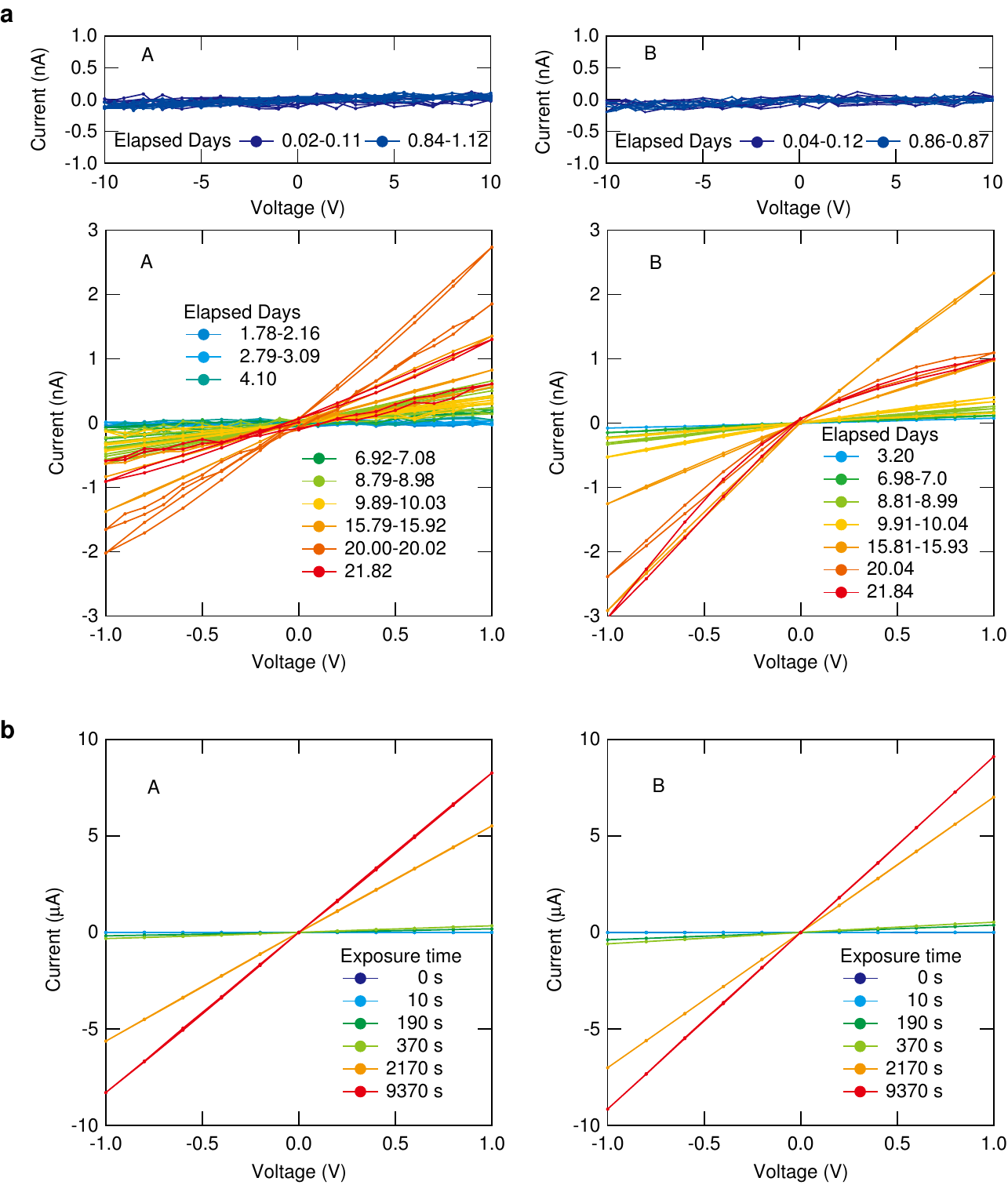}\\
\begin{flushleft}
{\footnotesize Figure S4: \textbf{Current-voltage ($I$-$V$) characteristics of hydrogen-terminated diamond kept in Ar gas and exposed to air.} \textbf{a}, $I$-$V$ characteristics of hydrogen-terminated diamond kept in the glove box after the vacuum transfer from a CVD chamber. The conductivity shown in Fig. 2a of the main text was obtained from a linear fit of these curves. The two different sets of curves correspond to the two different diagonal positions (A and B) on the diamond at which the prober needles were contacted; A and B correspond to the black and brown dots in Figs. 2a and 2b of the main text. The additional external purifier described in Methods of the main text was used for the first two days (1.88 days). \textbf{b}, $I$-$V$ characteristics of hydrogen-terminated diamond for cumulative air-exposure times. After measurements were made on the diamond that had been kept in the glove box (Fig. S4a), a test was conducted to see the effect of exposure to air. The diamond was exposed to air for a certain time and brought back into the glove box, where the $I$-$V$ characteristics were measured. This procedure was repeated while increasing the exposure time. The conductivity shown in Fig. 2b of the main text was obtained from a linear fit of these curves. Note the difference in the range of current (nA and $\mu$A) between \textbf{a} and \textbf{b}.}
\end{flushleft}
\end{figure}

\newpage

\begin{figure}
\includegraphics[width=15truecm]{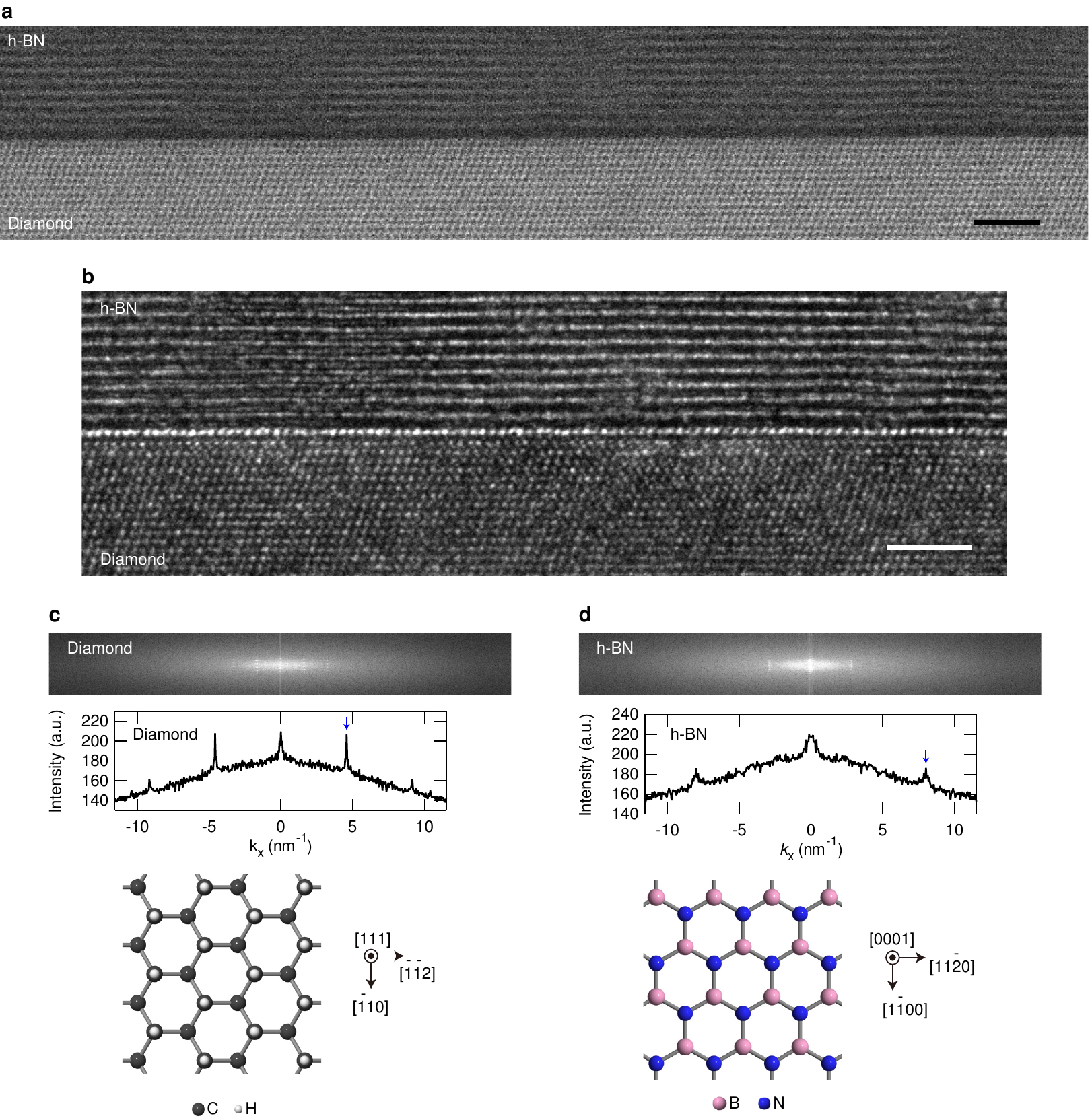}\\
\begin{flushleft}
{\footnotesize Figure S5: \textbf{STEM and HRTEM images of $h$-BN/hydrogen-terminated diamond heterostructure fabricated with air-free lamination process.} \textbf{a}, Scanning transmission electron microscope (STEM) image of a large-area $h$-BN/hydrogen-terminated (111) diamond heterostructure, showing a uniform spacing between the first $h$-BN layer and diamond over a large area. The scale bar is 2 nm. \textbf{b}, High-resolution TEM (HRTEM) image of the $h$-BN/diamond heterostructure. The scale bar is 2 nm. \textbf{c,d}, Fourier transform images obtained for the diamond and $h$-BN regions in the HRTEM image shown in (\textbf{b}) and their average intensity profiles along the lateral direction. The peak positions indicate that the $[1 \bar{1} 1 0]$ direction of $h$-BN is approximately aligned with the $[\bar{1} 1 0]$ direction of diamond in this specimen.}
\end{flushleft}
\end{figure}

\begin{figure}
\includegraphics[width=8truecm]{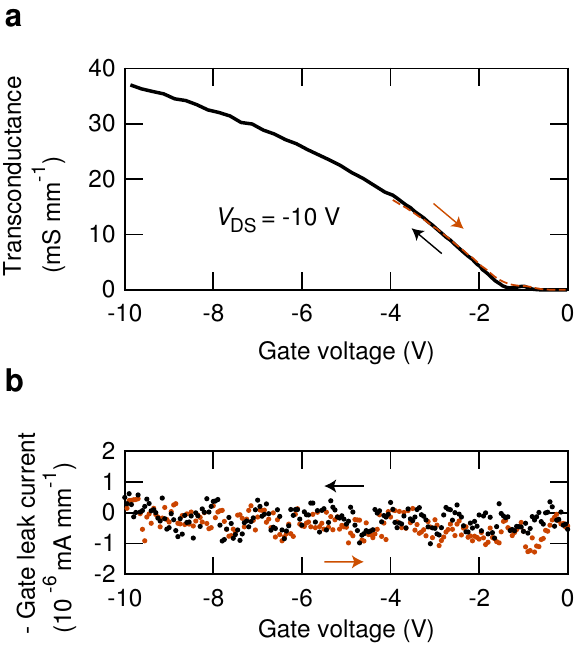}
\begin{flushleft}
{\footnotesize Figure S6: \textbf{Electrical characteristics of device C1 at 300 K.} \textbf{a}, Transconductance as a function of gate voltage for the drain voltage of -10 V. The solid and dashed lines correspond to gate sweeps from 0 to -10 V and from -4 to 0 V, respectively. \textbf{b}, Gate leak current as a function of gate voltage.}
\end{flushleft}
\end{figure}

\begin{figure}
\includegraphics[width=6truecm]{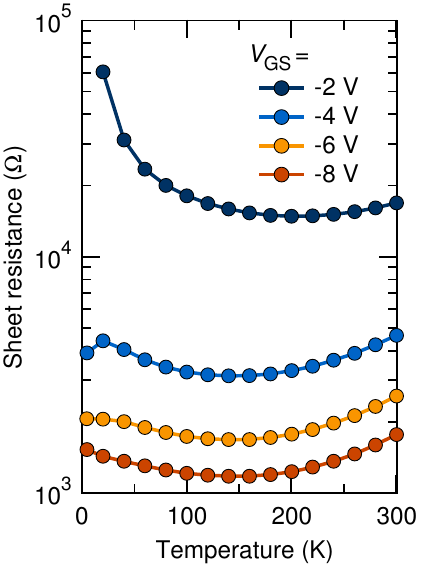}
\begin{flushleft}
{\footnotesize Figure S7: \textbf{Temperature dependence of sheet resistance of device C1 at different gate voltages.}}
\end{flushleft}
\end{figure}

\begin{figure}
\includegraphics[width=15truecm]{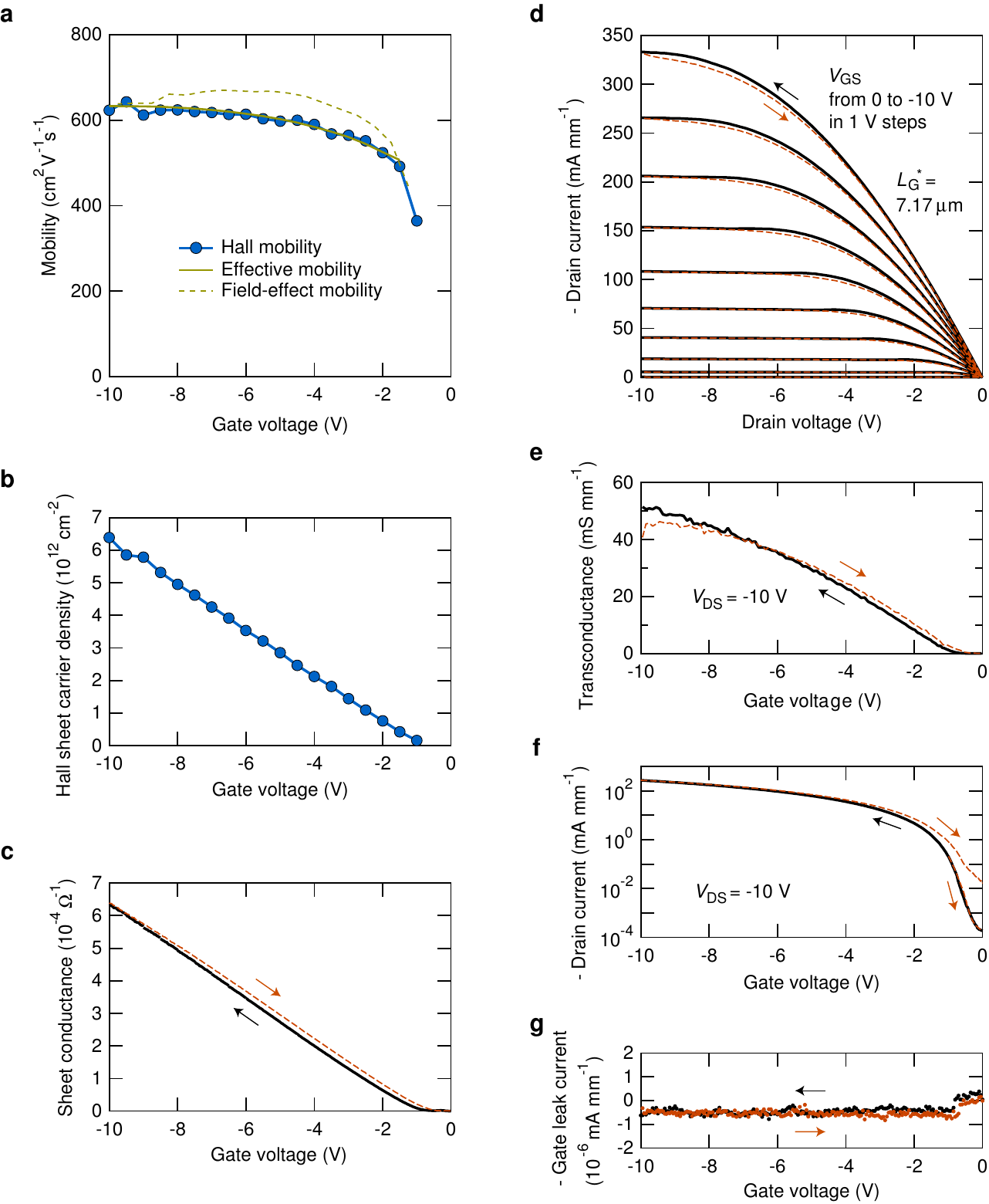}
\end{figure}

\begin{figure}
\begin{flushleft}
{\footnotesize Figure S8: \textbf{Electrical characteristics of device C2 at 300 K.} \textbf{a}, Hall, field-effect, and effective mobilities as a function of gate voltage. Dielectric constant of $h$-BN, $\epsilon_\mathrm{hBN}/\epsilon_0 = 3.0$, was used to calculate field-effect and effective mobilities. \textbf{b}, Hall carrier density as a function of gate voltage. \textbf{c}, Transfer characteristics in the linear region. The channel sheet conductance was measured with the four-probe configuration. The solid and dashed lines were obtained while the gate voltage was swept from 0 to -10 V and from -10 to 0 V, respectively. \textbf{d}, Output characteristics. The drain current measured with the two-probe configuration is plotted as a function of the drain voltage for different gate voltages. \textbf{e}, Transconductance as a function of gate voltage for the drain voltage of -10 V. The solid and dashed lines correspond to gate sweeps from 0 to -10 V and from -10 to 0 V, respectively. \textbf{f}, Drain current as a function of gate voltage at the drain voltage of -10 V. The solid line corresponds to a gate sweep from 0 to -10 V. The two dashed lines correspond to gate sweeps from -1 to 0 V and -10 to 0 V. \textbf{g}, Gate leak current as a function of gate voltage.}
\end{flushleft}
\end{figure}

\begin{figure}
\includegraphics[width=16truecm]{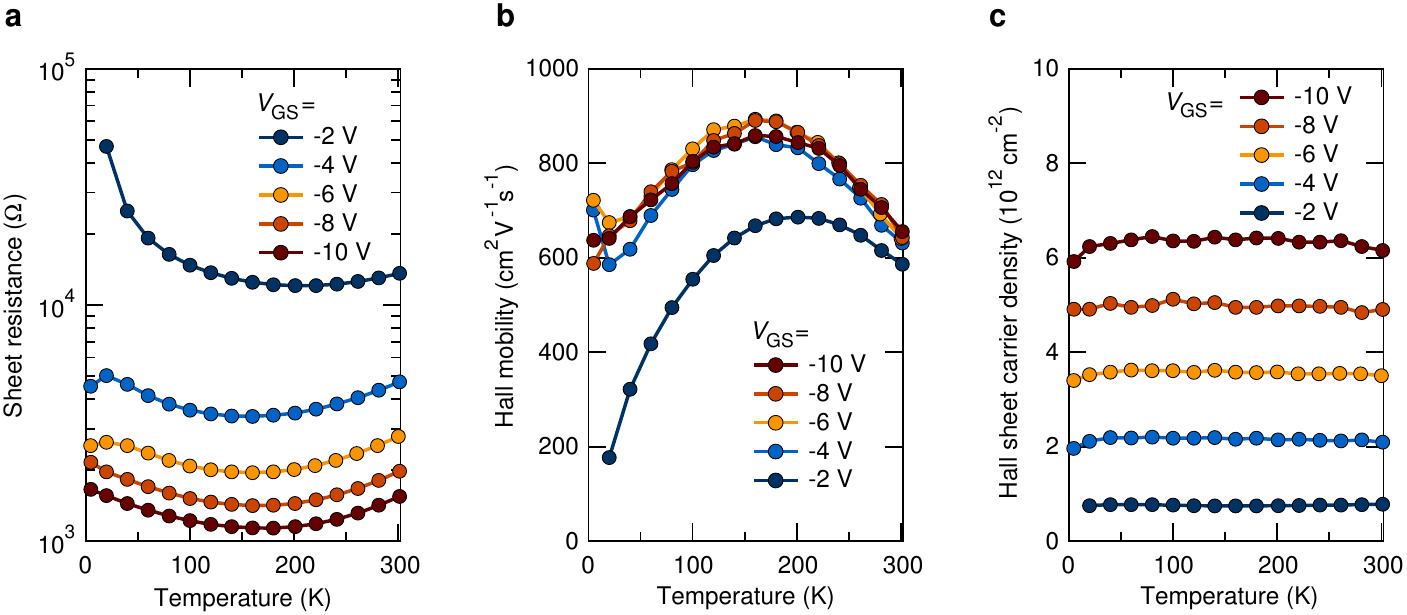}
\begin{flushleft}
{\footnotesize Figure S9: \textbf{Temperature-dependent electrical characteristics of device C2.} Temperature dependence of sheet resistance (\textbf{a}), Hall mobility (\textbf{b}) and Hall carrier density (\textbf{c}) at different gate voltages.}
\end{flushleft}
\end{figure}
\begin{figure}
\includegraphics[width=15truecm]{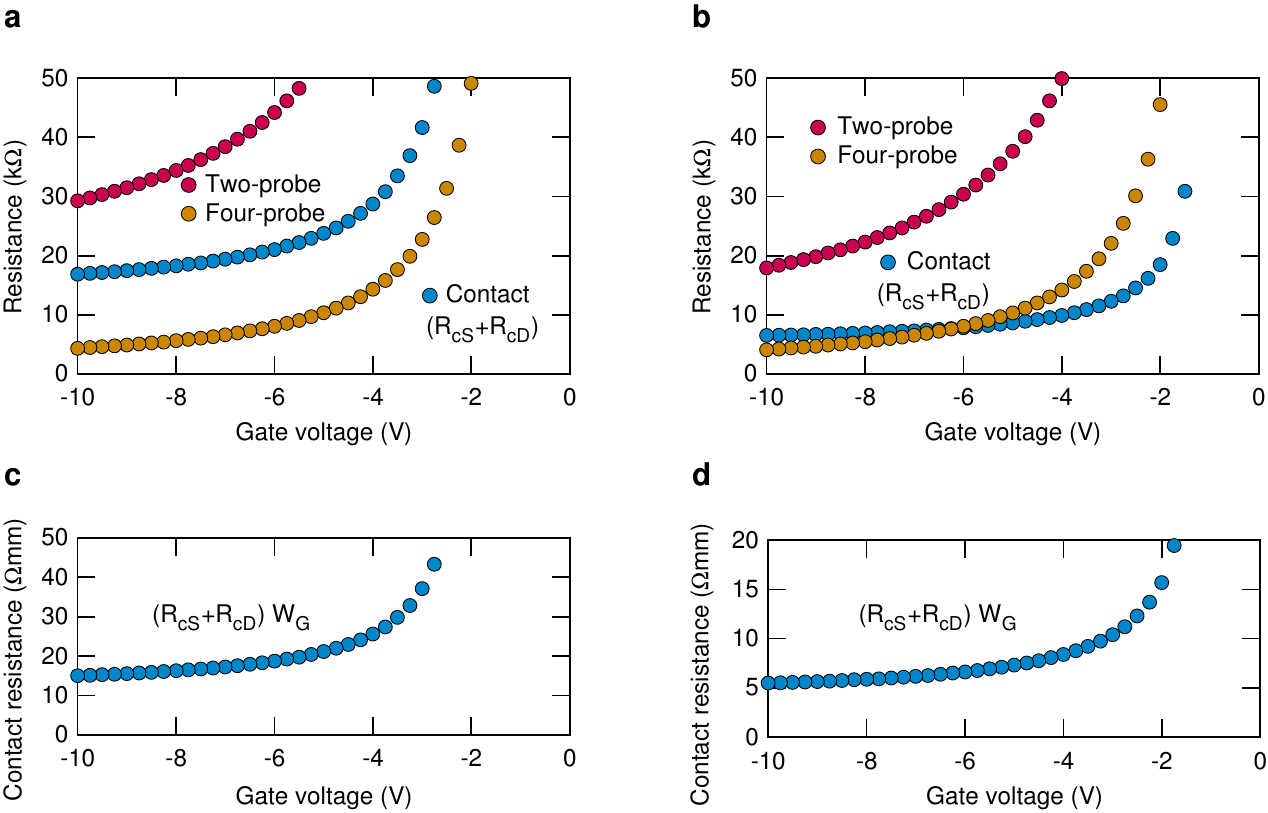}
\begin{flushleft}
{\footnotesize Figure S10: \textbf{Contact resistance of source and drain electrodes.} \textbf{a,b}, Gate voltage dependence of drain-source resistance ($R_\mathrm{2p}$) measured in a two-probe configuration, resistance measured in a four-probe configuration ($R_\mathrm{4p}$), and contact resistance ($R_\mathrm{cS}+R_\mathrm{cD}$) of the source and drain electrodes for devices C1 (\textbf{a}) and C2 (\textbf{b}). The contact resistance was calculated as $R_\mathrm{cS}+R_\mathrm{cD}=R_\mathrm{2p}-(L_\mathrm{G}^{*}/ L_\mathrm{p})R_\mathrm{4p}$. \textbf{c,d}, Contact resistance normalized by the channel width for devices C1 (\textbf{c}) and C2 (\textbf{d}).}
\end{flushleft}
\end{figure}

\begin{figure}
\includegraphics[width=14truecm]{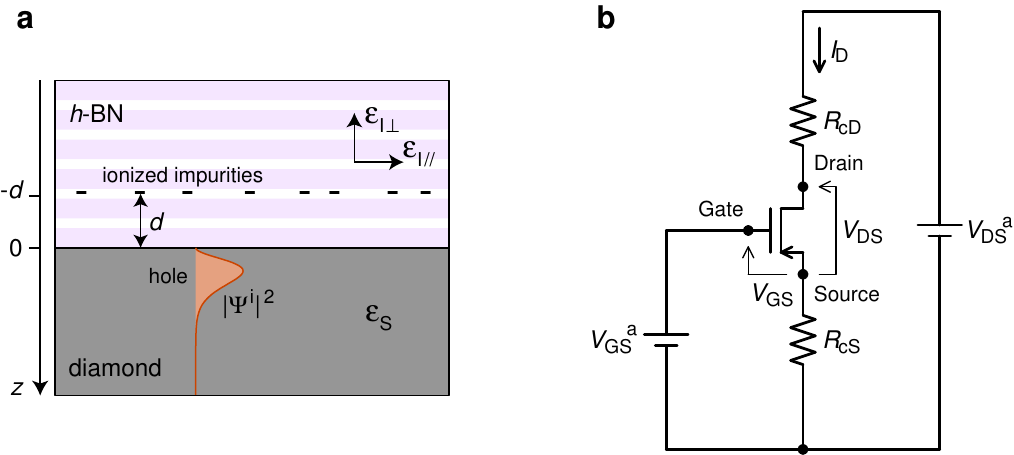}
\begin{flushleft}
{\footnotesize Figure S11: \textbf{Geometry for the calculation of mobility and circuit model for diamond FET.} \textbf{a}, Geometry for the calculation of mobility. The two-dimensional hole gas with the density distribution $\left|\Psi^{i}(z)\right|^{2}$ ($i =$ HH, LH, SO) is scattered by ionized impurities at a distance $d$ above the diamond surface. A general equation of the scattering rate for a distance $d$ is given in Supplementary Section C. For the calculation of mobility (Figs. S12a and S12b), we assumed $d=0$ because ionized impurities on the diamond surface or interface trapped charges are the most probable scattering source in our FET. The dielectric tensor of $h$-BN is assumed to be anisotropic. \textbf{b}, Circuit model for diamond FET including the effect of source and drain contact resistances.}
\end{flushleft}
\end{figure}

\begin{figure}
\includegraphics[width=12.7truecm]{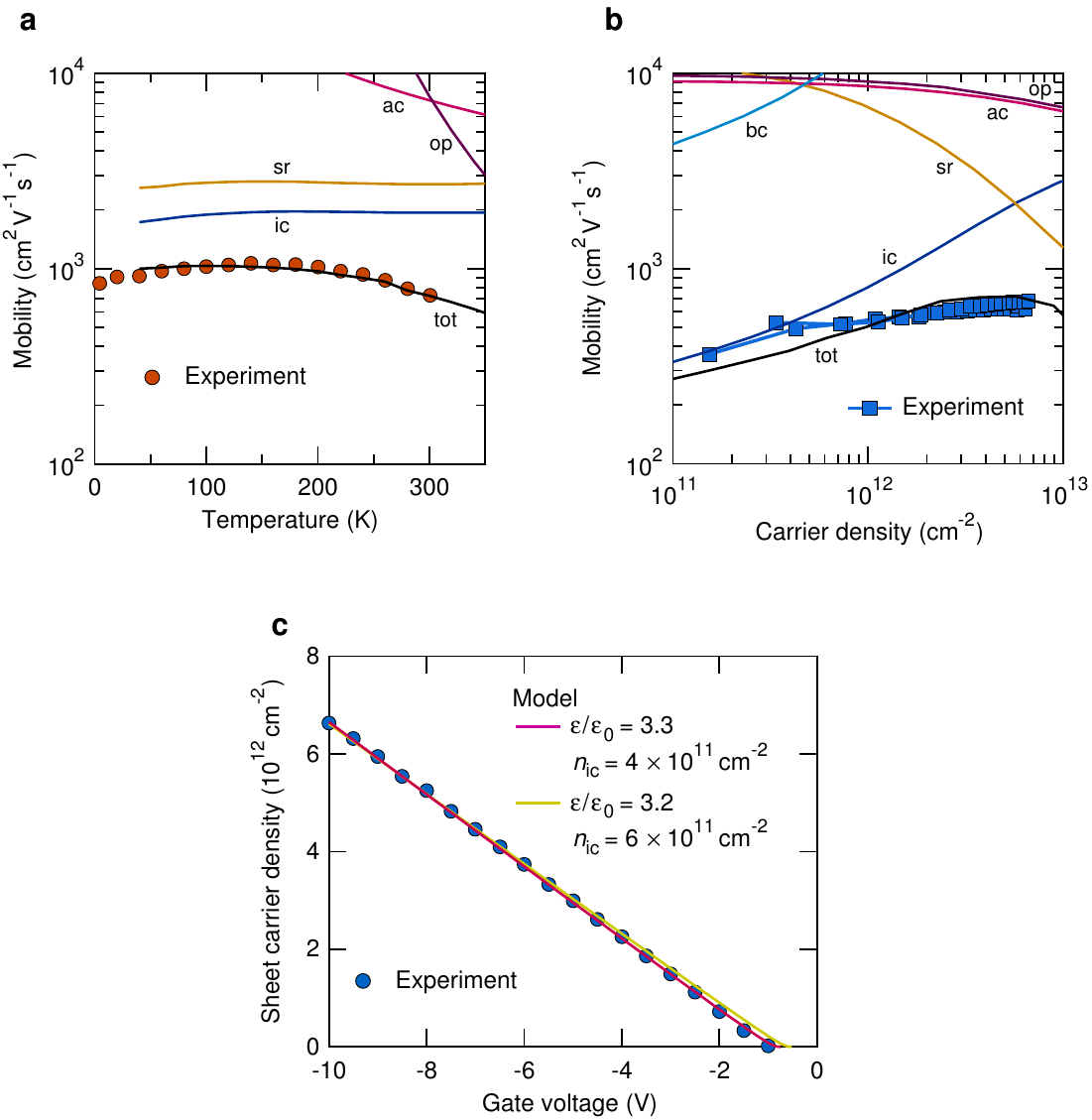}
\begin{flushleft}
{\footnotesize Figure S12: \textbf{Modelling of device characteristics.} \textbf{a}, Measured temperature-dependent Hall mobility of device C1 for a gate voltage of $-8$ V. The mobility calculated from the model described in Supplementary Section C is also shown. ``ic'', ``bc'', ``sr'', ``ac'', and ``op'' indicate the mobilities limited by interface charges, background charges (background ionized impurities), surface roughness, acoustic phonons, and non-polar optical phonons. ``tot'' indicates the mobility calculated with all the above scattering mechanisms taken into account. The mobility limited by background ionized impurities is higher than $5\times10^4$ cm$^2$V$^{-1}$s$^{-1}$ for the entire temperature range. \textbf{b}, Measured Hall mobility of devices C1 and C2 as a function of Hall carrier density. The mobility calculated from the model described in Supplementary Section C is also shown. \textbf{c}, Measured Hall carrier density of device C1 as a function of gate voltage at room temperature. The carrier density calculated from the model described in Supplementary Section D is also shown.}
\end{flushleft}
\end{figure}

\begin{figure}
\includegraphics[width=16truecm]{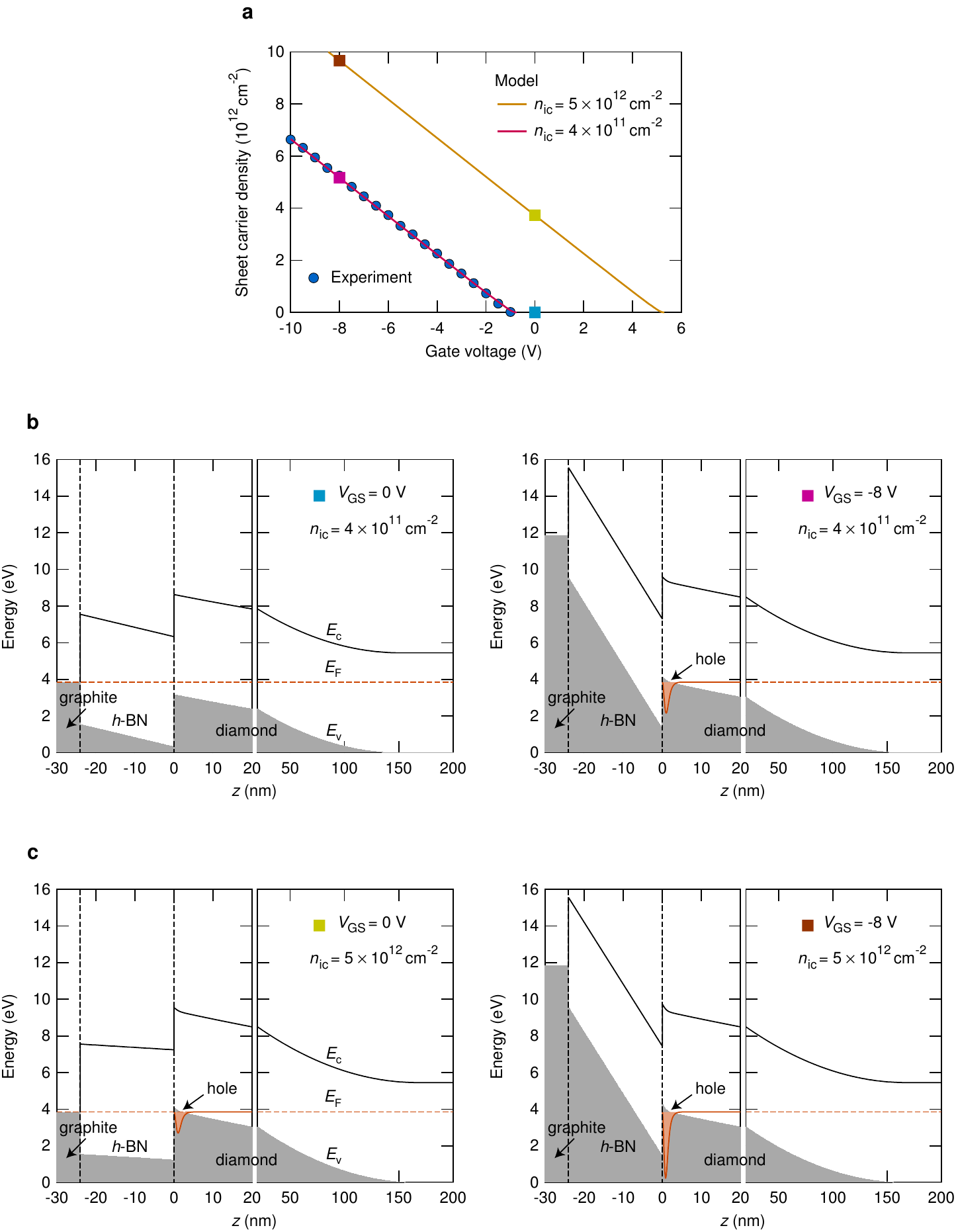}
\end{figure}

\begin{figure}
\begin{flushleft}
{\footnotesize Figure S13: \textbf{Energy-band diagrams for different densities of interface negative charges.} \textbf{a}, Gate-voltage dependence of carrier density calculated for different densities of interface negative charges: $n_\mathrm{ic} = 4\times10^{11}$ and $5\times10^{12}$ cm$^{-2}$. The measured Hall carrier density of device C1 is also shown. Four square points indicate four sets of carrier density and gate voltage for which energy-band diagrams shown in \textbf{b} and \textbf{c} were calculated. Normally-off and normally-on operations are indicated for $n_\mathrm{ic} = 4\times10^{11}$ and $5\times10^{12}$ cm$^{-2}$, respectively. \textbf{b}, Energy-band diagrams of the graphite (gate; $z{\textless}-24$ nm)/$h$-BN ($-24$ nm ${\textless} z {\textless} 0$)/diamond ($z{\textgreater}0$) heterostructure for $n_\mathrm{ic} = 4\times10^{11}$ cm$^{-2}$ under applied gate voltages of 0 and -8 V. \textbf{c}, Energy-band diagrams of the heterostructure for $n_\mathrm{ic} = 5\times10^{12}$ cm$^{-2}$ under applied gate voltages of 0 and -8 V.}
\end{flushleft}
\end{figure}

\end{document}